\documentclass[final,journal,twocolumn]{IEEEtran}

\newif\ifcmtr
\cmtrtrue

\ifcmtr 
\newcommand{\cmtr}[1]{ %
   [\color{red} \textbf{#1} \normalcolor]%
}%
\else
\newcommand{\cmtr}[1]{ %
}%
\fi

\ifcmtr 
\newcommand{\cmtrla}[1]{ %
   [\color{blue} \textbf{#1} \normalcolor]%
}%
\else
\newcommand{\cmtrla}[1]{ %
}%
\fi

\usepackage{hyperref}
\hypersetup{bookmarksnumbered=true}
\hypersetup{bookmarksopen=true}
\hypersetup{bookmarksopenlevel=2}
\hypersetup{colorlinks=true}
\hypersetup{linkcolor=black}
\hypersetup{citecolor=blue}
\hypersetup{filecolor=blue}
\hypersetup{menucolor=blue}
\hypersetup{urlcolor=blue}

\usepackage[]{siunitx}
\usepackage{mathptmx}
\usepackage{amsmath}
\usepackage{amssymb}
\usepackage{bm}
\usepackage{nicefrac}
\usepackage{upgreek}
\usepackage{listings} 

\usepackage{cleveref} 
\usepackage{graphicx}

\usepackage[ansinew]{inputenc}

\usepackage{tabularx}
\usepackage{threeparttable}
\usepackage{booktabs}
\AtBeginDocument{
\heavyrulewidth=.08em
\cmidrulewidth=.03em
\belowrulesep=.65ex
\belowbottomsep=0pt
\aboverulesep=.4ex
\abovetopsep=0pt
\cmidrulesep=\doublerulesep
\cmidrulekern=.5em
\defaultaddspace=.5em
}

\usepackage{doi}

\usepackage{color}
\definecolor{darkgreen}{rgb}{0,0.5,0}
\definecolor{darkblue}{rgb}{0,0,0.5}
\definecolor{brown}{rgb}{0.98,0.92,0.73}
\definecolor{red}{rgb}{1,0,0}
\definecolor{yellow}{rgb}{1,1,0}
\definecolor{blue}{rgb}{0,0,1}
\definecolor{green}{rgb}{0,1,0}
\definecolor{purple}{rgb}{1,0,1}
\definecolor{gray}{rgb}{0.8,0.8,0.8}
\definecolor{black}{rgb}{0,0,0}
\definecolor{white}{rgb}{1,1,1}
\definecolor{gold}{rgb}{1.,0.84,0.}

\usepackage{cite}
\usepackage{authblk}

\newcommand{\standout}[1]{\color{blue} #1}

\begin{document}

\title{Influence of the distribution of the properties of permanent magnets on the field homogeneity of magnet assemblies for mobile NMR}

\date{\today} 

\author{Y.P.Klein$^{1*}$}
\author{L.Abelmann$^{1,2}$}
\author{J.G.E.Gardeniers $^{1*}$}
\affil{
  $^1$ University of Twente, Enschede, The Netherlands\\
  $^2$ KIST Europe, Saarbr\"ucken, Germany\\
}

\maketitle 
\begin{abstract}
	We optimised the magnetic field homogeneity of two canonical designs
for mobile microfluidic NMR applications: two parallel magnets with an
air gap and a modified Halbach array.  Along with the influence of
the sample length, general design guidelines will be presented.  For a
fair comparison the sensitive length of the sample has been chosen to
be the same as the gap size between the magnets to ensure enough space
for the transmitting and receiving unit, as well as basic electric
shimming components.  Keeping the compactness of the final device in
mind, a box with an edge length 5 times the gap size has been defined,
in which the complete magnet configuration should fit.  With the
chosen boundary conditions, the simple parallel cuboid configuration
reaches the best homogeneity without active shimming
(0.5$\mathrm{B_{s}}$, 41 ppm), while the Pseudo-Halbach configuration
has the highest field strength (0.9$\mathrm{B_{s}}$, 994 ppm),
assuming perfect magnets.  However, permanent magnet configurations
suffer from imperfections, such as magnetisation, fabrication and
positioning errors, which results in worse magnetic field
homogeneities than expected from simulations using a fixed optimised
parameter set.  We present a sensitivity analysis for a magnetic cube
and the results of studies of the variations in the magnetisation and
angle of magnetisation of magnets purchased from different suppliers,
composed of different materials and coatings, and of different sizes.
We performed a detailed Monte Carlo simulation on the effect of the
measured distribution of magnetic properties on the mentioned
configurations.  The cuboid design shows a mean homogeneity of 430 ppm
(std dev. 350 ppm), the Pseudo-Halbach has a mean homogeneity of 1086
ppm (std dev. 8 ppm).

\end{abstract}
\section{Introduction}
\label {sec: Introduction}

Low-field and low-cost mobile microfluidic nuclear magnetic resonance
(NMR) sensors are very suitable for applications in chemical process
industry and in research, for example chemical analysis, biomedical
applications, and flow
measurements~\cite{Meribout2019, Zalesskiy2014,Mitchell2014,Danieli2010a,Lee2008,
  Kreyenschulte2015,Sorensen2015,Sorensen2014,Mozzhukhin2018}. The
design of a permanent magnet for an NMR sensor requires both a strong
magnetic field and a high field homogeneity within a defined region of
interest. In NMR, a high external magnetic field results in a
high spectral resolution and detection sensitivity.

However, field inhomogeneities compromise the spectral
resolution. Our aim with this research was to determine how the
distribution of the properties of permanent magnets affect the
magnetic field homogeneity of magnet configurations for mobile
NMR devices.

In the literature, several magnet shapes for mobile NMR sensors have been
reported. A broad overview of magnet developments up to 2009 can be
found in Demas et al.~\cite{Demas2009}. U-shaped
single-sided magnets~\cite{Bluemich1998,Meethan2014} and magnets with
specially shaped iron pole magnets~\cite{Marble2005} have been used
to explore surfaces. Mobile pseudo-Halbach
configurations~\cite{Vogel2016} and two cylindrical
magnets~\cite{SUN201313} have been applied for solid and liquid NMR
measurements. While the pseudo-Halbach generates a higher field,
ranging from \SI{0.7}{} to \SI{2.0}{T}~\cite{Danieli2010, Tayler2017, Moresi2003} compared to \SI{0.35}{} to \SI{0.6}{T} for the other configurations~\cite{Bluemich1998, Meethan2014, Marble2005, Lee2008}, the reported field homogeneities without electric shimming seem to be independent of the design, ranging from \SI{20}{ppm} to \SI{606}{ppm}~\cite{SUN201313, Lee2008, Sahebjavaher2010, Moresi2003, Meribout2019, Chen20073555}.  
Comparing the two most reported mobile liquid NMR sensors, it further stands out that there is no obvious relation between the size of the sensor and the choice of the magnet configuration.

To achieve more insight into possible guidelines for the magnet design, in this paper
a modelling study will be presented from which the homogeneity and
field strength at specific locations in the gap of the magnet
configuration is derived numerically.  It is widely experienced that
after building such a permanent magnet configuration, the homogeneity
reached in practice does not exhibit the same results as in the
simulation~\cite{Danieli2010,Moresi2003,Horton1996,Soltner2010, Ambrisi20102747}, which
can be caused by several factors. The magnetisation of permanent
magnets depends highly on the temperature, as well as on the remanent
magnetisation~\cite{Haavisto2011}. This remanent magnetisation can
change over time due to shock-induced
demagnetisation~\cite{Li2013,Royce1966}, external magnetic
fields~\cite{Lee2011}, a degrading of the magnetic material caused by
oxidation~\cite{Li2003a}, as well as broken or chipped off pieces (since magnets
are very brittle)~\cite{Horton1996}. Next to material related
differences, fabrication inaccuracies such as variations in the dimensions and
magnetisation angles affect the field created by
a permanent magnet. On top of that, magnet configurations can never be
assembled perfectly. Errors in placement may induce a tilt or an axial
offset of the magnet.

\begin{figure} 
	\centering
	\includegraphics[width=\linewidth]{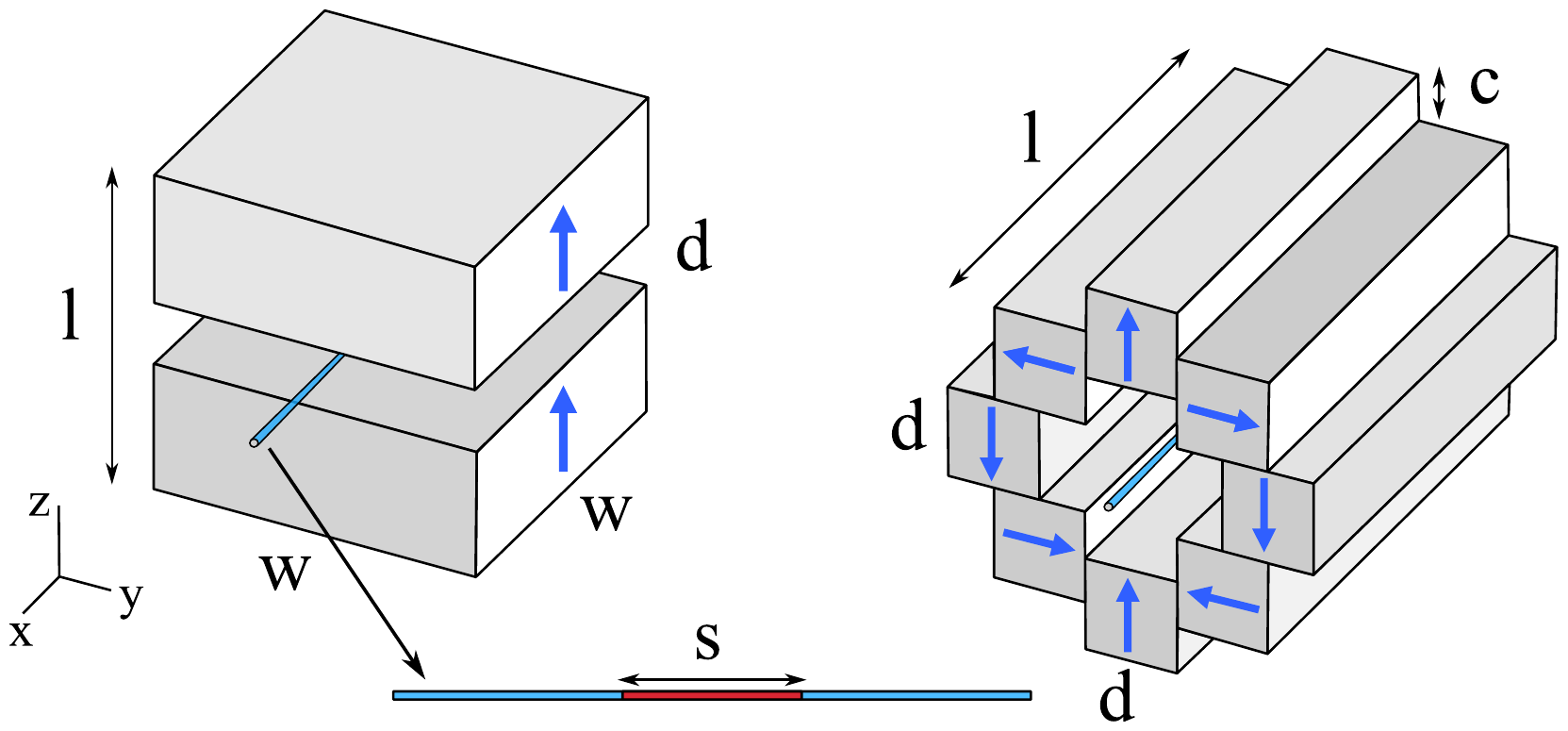}
	\caption{Magnet configurations for microfluidic NMR. The
          sample under investigation is inside a tubular channel
          running through the center of the configurations, and has a
          length $s$. The arrows indicate the magnetisation of each
          individual magnet. Left: Cuboid configuration. The height of
          the stack $l$ is fixed at five times the gap between the
          magnets $d$. The width $w$ of the stack is optimised for
          minimum field inhomogeneity over the sample length $s$.
          Right: Pseudo-Halbach configuration. Again $l$ is fixed to
          $5d$, but now the magnet recess $c$ is optimised. }
	\label{fig1}
\end{figure}

We carried out an extensive numerical sensitivity analysis of a single
cubic magnet using these variations. We measured the
variations in the magnetisation and magnetisation angle of magnets composed of
different materials, with different coatings, and with different sizes, obtained from
different manufacturers. The two main magnet configurations
investigated are a system of two parallel magnets and a Pseudo-Halbach
configuration~\cite{Demas2009}, shown in Fig.~\ref{fig1}. One
configuration of each type has been designed and optimised for the
following boundary conditions. The sensitive length of the channel
(\textit{s}) has been chosen to be the same as the gap size
(\textit{d}). For example: In case a maximal magnet size of
\SI{50x50x50}{mm} is required, the gap size turns out to be
\SI{10}{mm}. All dimension specifications are scalable and will be
normalised by the gap length. Scaling the dimensions bigger or
smaller will result in an increased or decreased sample length relative
to the dimensions of the gap, while the magnetic field properties
within the region of interest will stay the same. The magnetic field
has been normalised to the residual magnetic flux density $B_\text{s}$ (T) of the used
magnetic material. The cuboid configuration consists of two cuboid
magnets with a height of $2d$ and a width of \num{4.72}$d$. The
Pseudo-Halbach configuration consists of eight bar magnets, each
with the dimensions $d \times d \times 5d$. The measured variations in the magnets
have been used to perform a Monte Carlo simulation to
provide insight into how the homogeneity of those configurations varies
after assembling. The results have been verified with field
measurements done with a Tesla meter.  The sample channel in most
published microfluidic NMR sensors has a high ratio of sample length over
inner diameter ($s/d_\text{i}$) (\SI{5.0}{} over \SI{0.4}{mm}
in~\cite{Gardeniers2009}, \SI{30}{} over \SI{1.0}{mm}
in~\cite{Kalfe2015}, and \SI{2.9}{} over \SI{0.15}{mm}
in~\cite{McDonnell2005}). Therefore we focus on a high
field homogeneity in mainly one dimension ($x$-axis).

\section{Methods}
\label {sec: Methods}

\subsection{Determination of variation in magnet properties}	
The variations in the properties of the magnets have been measured with a 3D Hall-probe (THM1176 Three-axis Hall Magnetometer, Metrolab).
The setup for the configuration measurements contains a stable temperature environment (\SI{38.0(5)}{\celsius}) and a Hall sensor from Projekt Elektronik GmbH (Teslameter 3002/Transverse Probe T3-1,4-5,0-70) in combination with a motorised linear stage.
Since the sensor is in a fixed position and only the magnet was moved for the measurement, field variations within the oven have no influence on the measurement.
Different kinds of magnets have been purchased.
We chose different materials, coatings, sizes and manufacturers, shown in Table~\ref{tab10}.

\begin{table*}
	\centering
  \caption{Purchased permanent magnets. \label{tab10}}
  \begin{threeparttable} 
    \begin{tabularx}{0.9\textwidth}{l@{\extracolsep{\fill}}llllll}
      \hline
      \hline
      Manufacturer & Dimension & Material & $B_\text{r}$ & $BH_\text{max}$  & Coating &
      Abbreviation  \\
      & (mm) & &(mT) & (kJ/m$^3$) & & \\
      \hline
      Supermagnete & 45$\times$30  \tnote{1} &
      NdFeB (N45)& 1320-1360 & 340-372 
      & Ni-Cu-Ni & Su45Nd45NCN\\
      Supermagnete & 7$\times$7$\times$7 & NdFeB (N42) & 1290-1330 & 318-342 
      & Ni-Cu-Ni & Su7Nd42NCN \\
      Supermagnete & 7$\times$7$\times$7 & NdFeB (N42) & 1290-1330 & 318-342  & Ni-Cu & Su7Nd42NC \\
      HKCM & 7$\times$7$\times$7 & NdFeB (N35)& 1180-1230 & 263-287 
      & Ni & HK7Nd35N \\
      HKCM & 7$\times$7$\times$7 & Sm2Co17 (YXG28) & 1030-1080 & 207-220 
      & Ni & HK7Sm28N \\
      Schallenkammer Magnetsyteme & 7$\times$7$\times$7 & Sm2Co17 (YXG26H) & 1020-1050 & 191-207 
      & - &  Sc7Sm26\\
      \hline
      \hline
    \end{tabularx}
    \begin{tablenotes}\footnotesize 
    \item[1] diameter$\times$height, axially magnetised
    \end{tablenotes}
  \end{threeparttable}
\end{table*}

\subsection{Stray field calculation}
Calculations of the magnetic stray fields were performed using CADES
simulation software,  described by Delinchant et
al.~\cite{Delinchant2007}. The magnetic interactions are modelled with
the MacMMems tool, which uses the Coulombian equivalent charge method
to generate a semi-analytic model. 

\begin{eqnarray}
\textbf{B}(\textbf{r})=\iint
  \limits\dfrac{\sigma (\textbf{r}-\textbf{r}^\prime)}{\lvert\textbf{r}-\textbf{r}^\prime\rvert^{3}}
  ds \text{, } \sigma=\mu_0\textbf{M}\cdot\textbf{n}  \nonumber 
\end{eqnarray}

\noindent Here, $\textbf{B}$ is the magnetic field (T) and
$\textbf{M}$ the magnetisation of the permanent magnet (\SI{}{A/m}),
$\textbf{r}$ and $\textbf{r}^\prime$ define the observation point and
its distance to the elementary field source area $ds$. The integral is
taken over the surface of the magnets. $\sigma$ (T) is
the magnetic surface charge, and $\textbf{n}$ the unit vector normal to the
surface.

The CADES framework, including a component generator, component calculator, and component optimiser, generated the final equations, which are used to calculate and optimise the designs.

\subsection{ Design optimisation procedure}
\label{sec:optimisationProcedure}
The stray field calculations are used to optimize particular magnet
configurations with respect to the inhomogeneity of the magnetic field
over the length of the sample. This inhomogeniety is captured in a
single valued metric defined as
the root mean square of the difference between the
\textit{z}-component of the mean field
$B_\text{mean}$ and the \textit{z}-component  of field along the
sample $B_z$, averaged of the sample length $s$ and related to the
mean field:

\begin{align}
  \frac{1}{sB_\text{mean}}
  \int_{-s/2}^{s/2}\sqrt{(B_z-B_\text{mean})^{2}}dx \nonumber
\end{align}

\begin{figure}  
  \centering
  \includegraphics[width=0.9\linewidth]{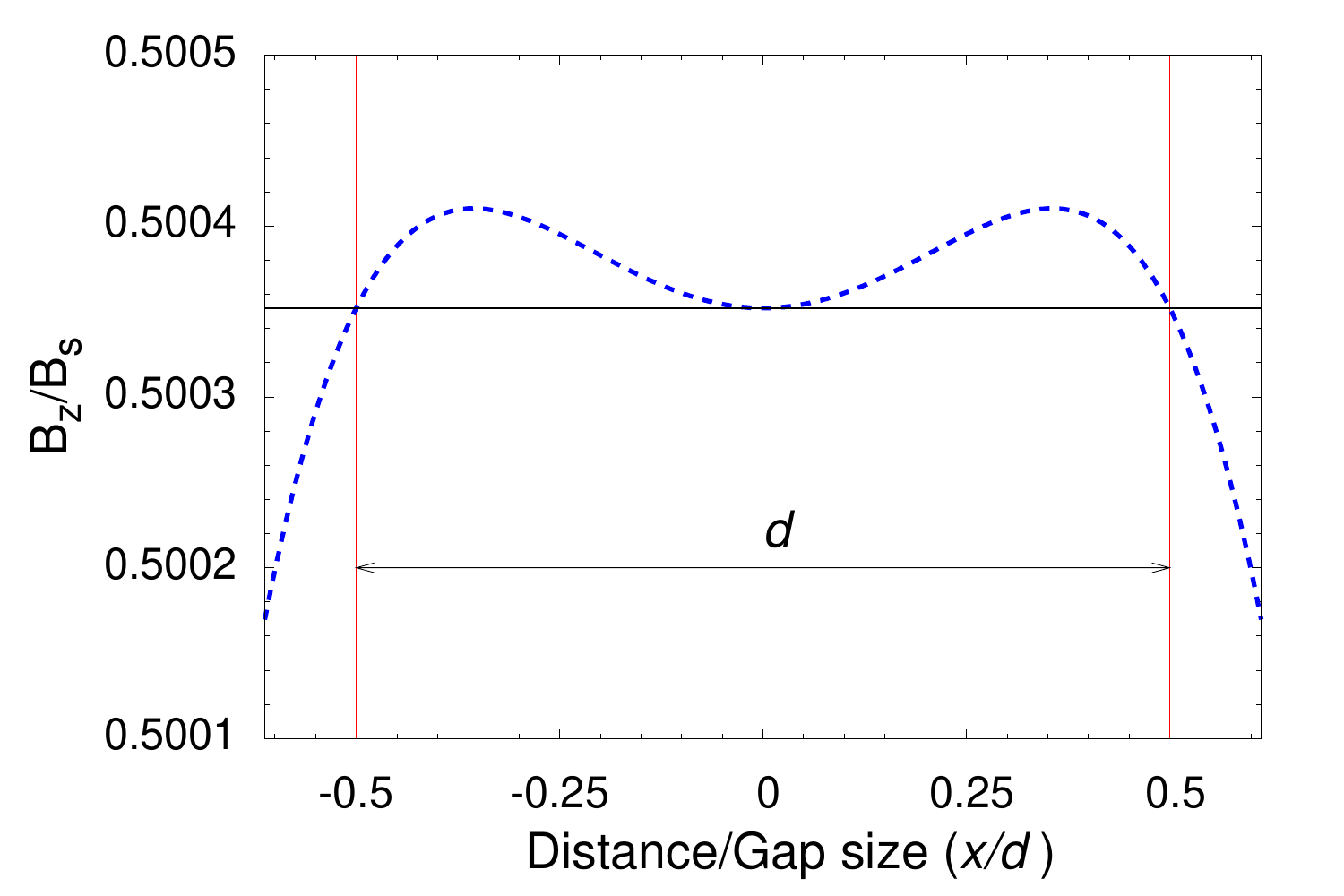}
  \caption 
  {Field profile of the cuboid configuration as a function of the
    relative distance from the centre between the magnets. In the
    optimised situation for a sample length equal to the gap size, the
    field in the centre equals the field at the edges ($x$ =
    $\pm 0.5d$).}
  \label{fig2}
\end{figure}

Minimisation of this metric leads to the simple rule that the field at
the edges of the sample should equal the field in the center. We
illustrate this for the cuboid configuration, illustrated in
Fig.~\ref{fig1}. Fig.~\ref{fig2} shows the magnetic field along the
sample of the optimised cuboid configuration, in which the field is
the same in the centre and at the edge of a sample. The field is
symmetric, showing a valley in the middle and two peaks in the
directions of the edges. After those maxima, the field decreases with
the distance to the centre.

\begin{figure} 
  \centering
  \includegraphics[width=0.9\linewidth]{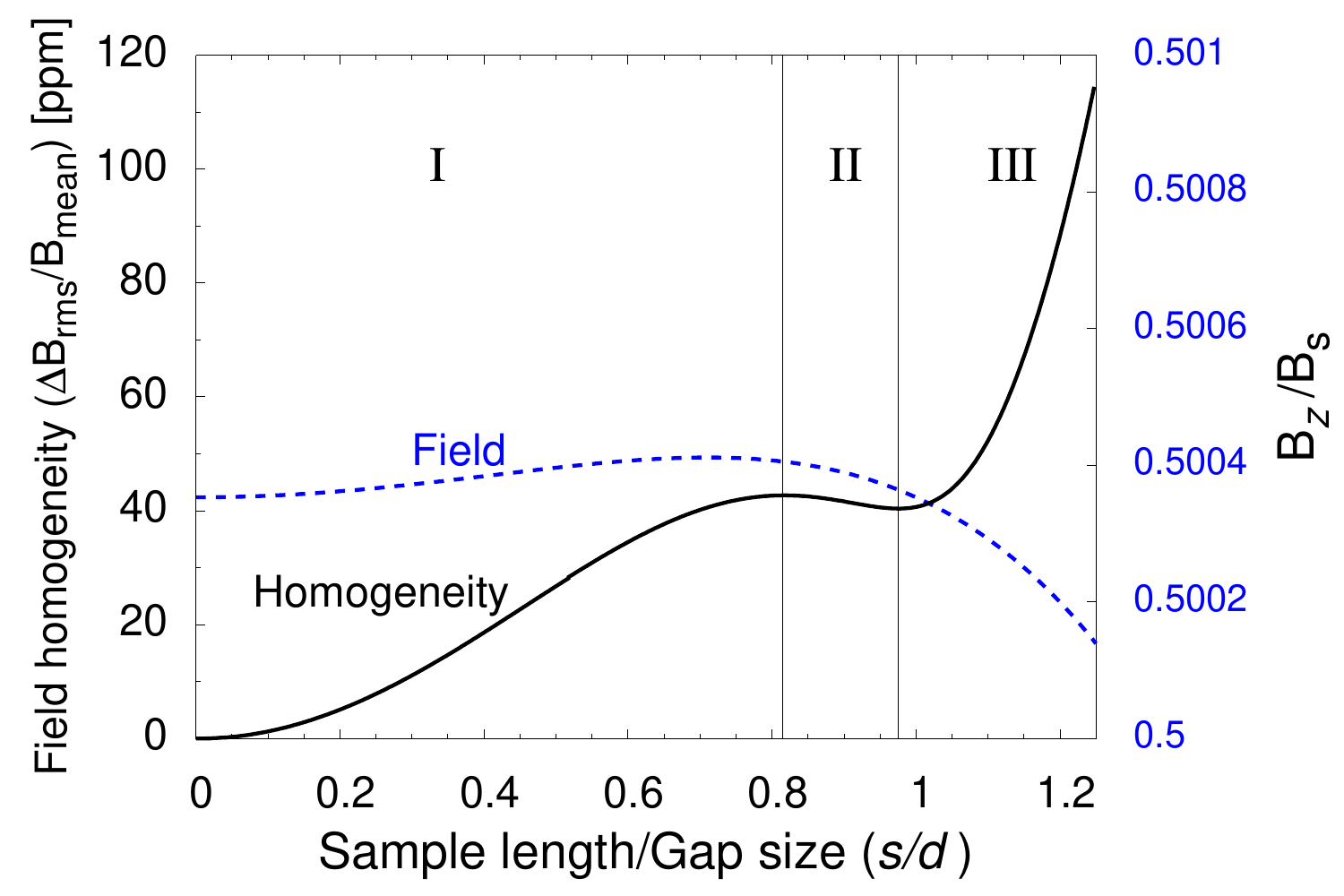}
  \caption{Field inhomogeneity and field as a function of the sample
    length for the cuboid configuration. The homogeneity has been
    optimised for a sample length equal to the gap size. With increasing
    sample length, both the field and the field inhomogeneity increase
    theoretically (region I). The field reaches a local maximum of
    \num{0.50041}$B_\text{s}$ at a distance of \num{0.71}$d$ from the
    centre. Above this distance the homogeneity of the sample stays
    approximately the same (region II). When the sample length increases
    more than the gap size, the inhomogeneity strongly increases
    (region III). If we choose the field at the edge of the sample to
    be equal to the centre of the sample, we are very close to the
    minimum homogeneity in region II.}
  \label{fig3}
\end{figure}

Fig.~\ref{fig3} shows how the field homogeneity develops with
increasing sample length while keeping the previously optimised
parameter set constant. Three regions can be seen. In the first one
the field increases from \num{0.50035}$B_\text{s}$ to
\num{0.50041}$B_\text{s}$, which means that the minimum field of
\num{0.50035}$B_\text{s}$ stays the same while the maximum field is
increasing until it reaches its global maximum, hence the
inhomogeneity is also increasing. In the second region the
inhomogeneity stays almost constant. In the third region the field
decreases below the previous minimum, which results in a drastic
increase of the inhomogeneity. Therefore, the lowest inhomogeneity
between two points can either be reached by keeping the sample as
short as possible or when the field at the sample edges is
approximately equal to the field at the center. Since the signal in
NMR is proportional to the sample volume, we optimise for the latter
condition.

\section{Results and Discussion}
\label {sec: Results}

The field uniformity of the various designs is determined by the
design itself as well as the manufacturability. One major point of
concern is the variation in the value and alignment of the
magnetic moment of the permanent magnets. Therefore we first present
the distribution of these properties for a range of commercial
magnets. We subsequently optimise the designs with respect to uniformity and
analyse their sensitivity to magnet variation using sensitivity
matrices and Monte-Carlo
simulations. These model predictions are than compared with six
realisations of the different designs.

\subsection{Variation of properties of commercial permanent magnets}
We measured the variations in the magnetisation and magnetisation
angle of magnets obtained from different companies (Supermagnete, HKCM
and Schallenkammer Magnetsysteme), compositions (NdFeB N45, NdFeB N42,
Sm2Co17 YXG28, Sm2Co17 YXG26H), coatings (Ni-Cu-Ni, Ni-Cu, Ni, no
coating), and sizes (cylinders with a diameter of \SI{45}{} and height
of \SI{30}{mm} or cubes of \SI{7x7x7}{mm}). Of each set, \num{50}
magnets were analysed. An overview of the distributions in residual
magnetic flux density and angle of magnetisation is given in
Table~\ref{tab3}. The raw data is provided in the supplementary
material (Appendix \ref{sec:appendixMagnetDistribution}).


\begin{table}
  \centering
  \caption{Measured coefficient of variation (CV) of residual magnetic flux density 
    and standard deviations of magnetisation angle of magnets with different materials,
    coatings, sizes and manufacturers.\label{tab3}}
  \begin{threeparttable} 
  \begin{tabular*}{0.8\columnwidth}{l@{\extracolsep{\fill}}SS}
    \hline
    \hline
    Magnet &  {$B_\text{mean} , CV[\si{\percent}]$}
    & { $ \phi $ [\si{\degree}] }\\
    \noalign{\vskip 0.5mm} 
    \hline
    Su45Nd45NCN & 0.7(3) & 0.0(1) \tnote{1}\\ 
    Su7Nd42NCN   & 0.8(2) & 0.7(2)\\ 
    Su7Nd42NC     & 0.6(3) & 0.0(1)\\ 
    HK7Nd35N       & 0.3(3) & 0.4(2)\\ 
    HK7Sm28N      & 1.0(3) & 0.2(1)\\ 
    Sc7Sm26          & 1.6(2) & 1.0(2)\\
    \hline
    \hline
  \end{tabular*}
  \begin{tablenotes}\footnotesize 
    \item[1] The values between brackets are the absolute
  standard errors of the last shown digits.
    \end{tablenotes}
  \end{threeparttable}
\end{table}

On average, the residual flux density varies by 1\% of $B_\text{mean}$. The
cylindrical magnet, which has a more than \SI{50}{} times higher
magnetic volume than the cubes, shows roughly the same variation in
magnetisation. From this, we can conclude that inaccuracies in the
dimensions are not the main cause of the variation in the
magnetisation. The uncoated Sm2Co17 shows a higher variation in
magnetisation than the coated magnets, which could be caused by
oxidation or small damage to the magnet since unprotected sharp edges
of magnets tend to break off easily. Different coatings do not show a
clear trend regarding the magnetisation standard variation or the
variation in the magnetisation angle. The offset angle varies on
average by less than \SI{1}{\degree}. There is no clear relation
between the variation in magnetisation strength or orientation and
material, coating or manufacturer.


\subsection{Design optimisation}

The optimisation method described in
section~\ref{sec:optimisationProcedure} was applied to both the cuboid
and the pseudo-Halbach design.

\paragraph{Optimisation of the cuboid configuration}
The cuboid configuration consists of two parallel cuboid magnets. The
length $L$ of the entire configuration has been chosen to be five times
the gap size $d$. The width $W$ was used to tune the field in between
the magnets. The optimisation procedure aims to find a width
for which the field in the centre and at the sample edge is the same.

\begin{figure} 
  \centering
  \includegraphics[width=0.9\linewidth]{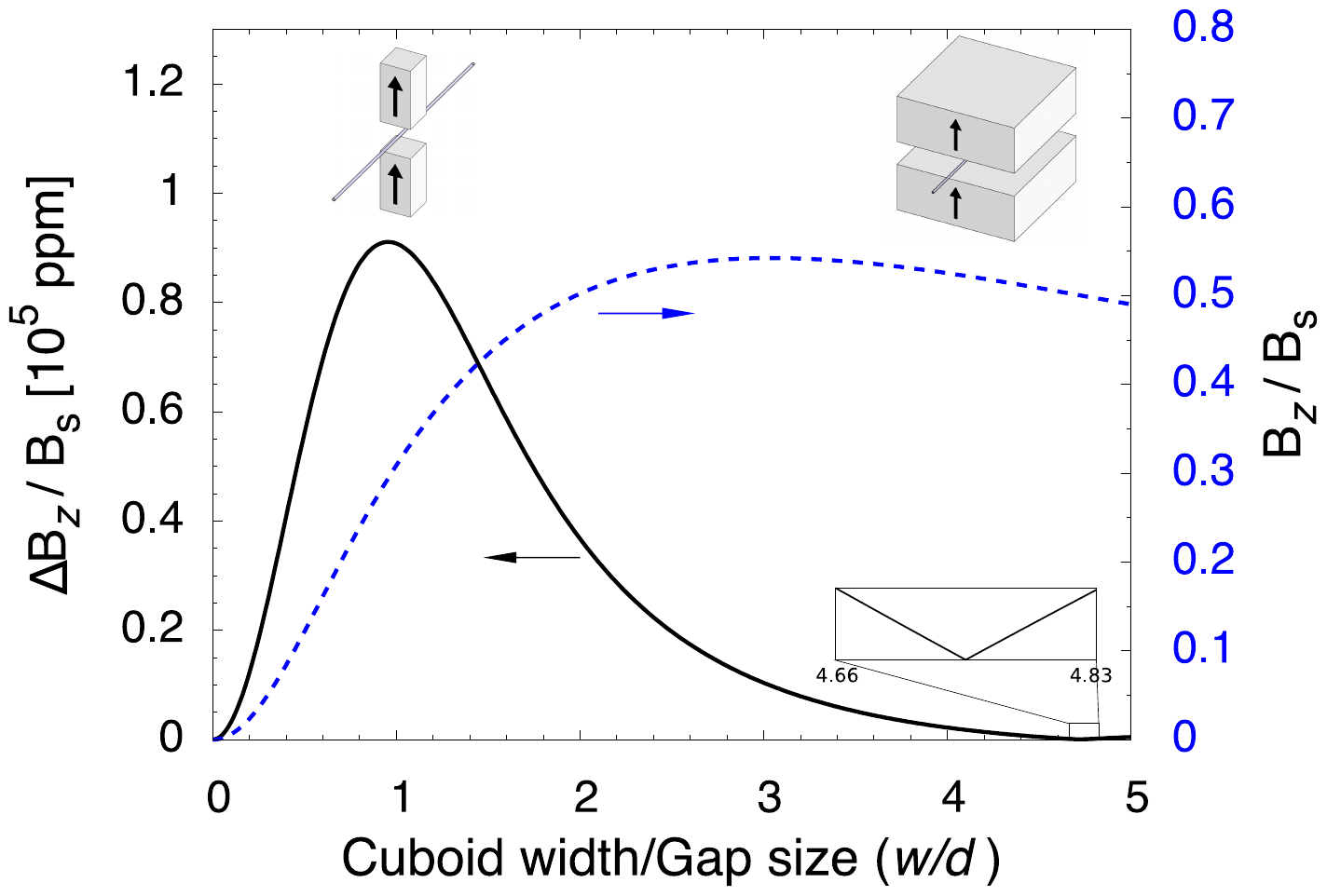}
  \caption
  {Field strength in the middle of the configuration and
    difference of the field centre and the edge of the sample,
    both as functions of the ratio of the cuboid width over the length. The
    field increases up to \num{0.54}$B_\text{s}$ at a cuboid width
    of \num{3.0375} times the gap size. The inset shows the
    field difference dropping to zero at a width/gap ratio of
    \num{4.72}.}
  \label{fig4}
\end{figure}

Fig.~\ref{fig4} shows that the magnetic field in the centre increases 
to its maximum of \num{0.54}$B_\text{s}$ at a width of
\num{3.0375}$d$. Increasing the width further results in a reduction
of the magnetic field, caused by the larger distance from the 
edges of the magnet to the centre. The difference between the magnetic field in the
centre and that at the sample edge increases until it reaches a maximum, when the
width equals the gap size. From this point the difference decreases
until it reaches a minimum at a width/gap ratio of \num{4.72}. The stray field at a
distance equal to the gap size is \num{0.24}$B_\text{s}$. 

\paragraph{Pseudo-Halbach}
The pseudo-Halbach configuration consists of
eight magnets, arranged in such a way that the field in the bore is
enhanced while the external stray field is minimised. The magnets have
a fixed dimension $d \times d \times 5d$. To tune the homogeneity,
the position of the magnets in the corners is fixed, while the other
magnets are spread out over a distance $c$  (Fig.~\ref{fig1}). The
width starts at $w = 3d$ to ensure a minimum bore width $d$ and ends
at $w= l$, due to the previously chosen boundary conditions.  

\begin{figure}
  \centering
  \includegraphics[width=0.9\linewidth]{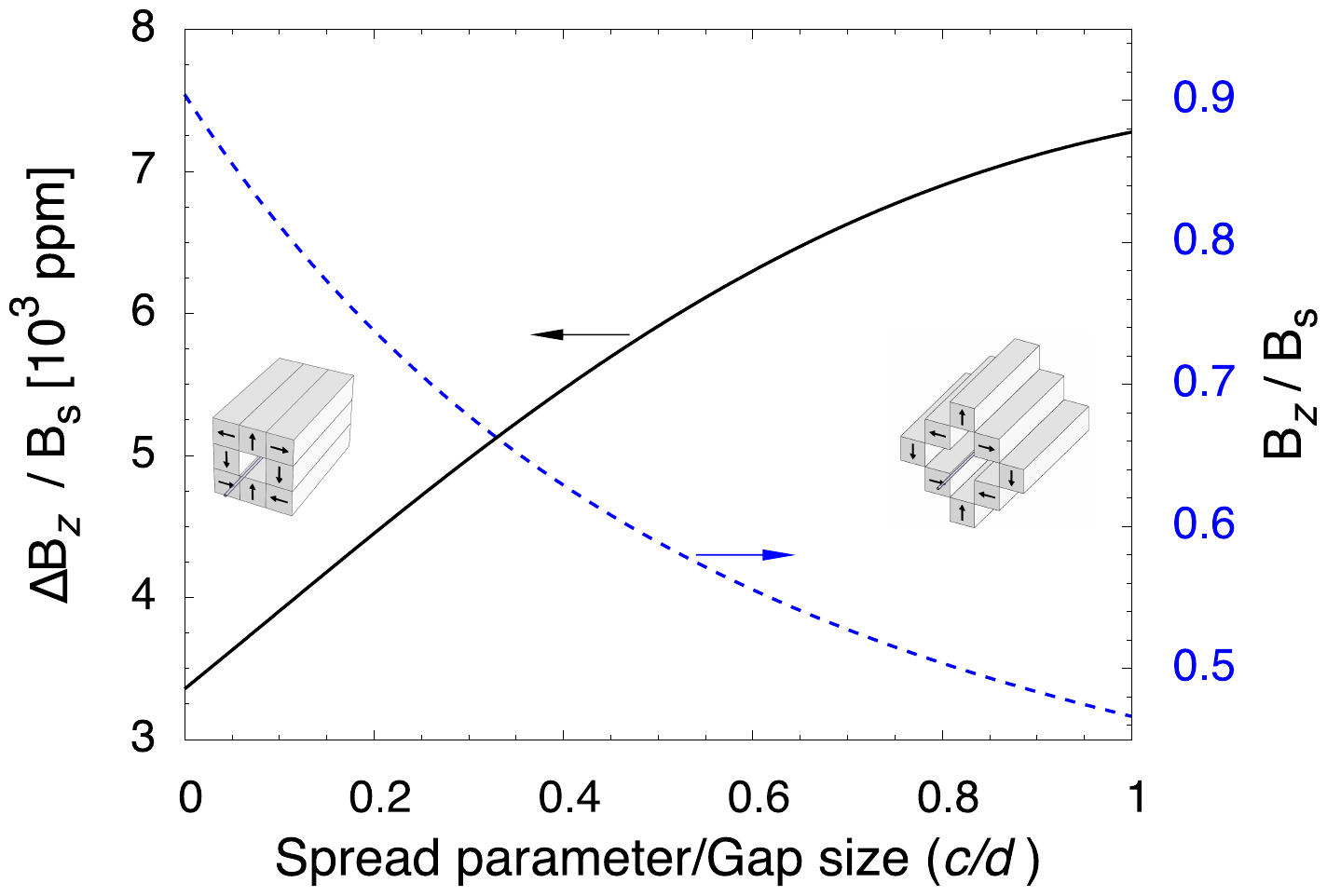}
  \caption{Spreading the middle magnets has been used to change
    the normalised field strength and field difference. At a
    spread parameter $c = 0$, a minimal field difference of
    \SI{3365}{ppm} and a field strength of \num{0.90}$B_\text{s}$
    can be reached.}
  \label{fig5}
  \vspace{4ex}
\end{figure}

Spreading the configuration increases the distance of the middle
magnets, which produces a decreased magnetic field strength
(Fig.~\ref{fig5}). With this configuration the convex field profile
has no chance to change to a concave profile. Therefore a minimum can
not be reached. With the most compact magnet arrangement
($c$=0), a field of \num{0.9}$B_\text{s}$ and a field
difference of \SI{3365}{ppm} can be achieved. The stray field at a
distance equal to the gap size from the surface is
\num{0.07}$B_\text{s}$.

In table~\ref{tab1} the major specifications of the two optimised
configurations are compared. The pseudo-Halbach configuration achieves
\num{0.9}$B_\text{s}$, a \num{1.8} times higher field than the Cuboid
configuration, while the stray field at a distance of $d$ from the
magnet surface is \num{0.07}$B_\text{s}$, which is \num{3.4} times
lower.  In terms of homogeneity, the Cuboid configuration achieves a
homogeneity of \SI{41}{ppm}, which, compared to the pseudo-Halbach
configuration, is \num{24.2} times better.

\begin{table}
  \centering
  \caption{Comparison of magnetic properties of different magnet configurations.}
  \begin{tabular}{lSSS}
    \hline
    \hline
    & {$B_\text{max}$}  & {$B_\text{stray}$} & {$\Delta B_\text{rms}/B_\text{mean}$} \\
    & {$[B_\text{s}]$} & {$[B_\text{s}]$} & {$[$ppm$]$} \\
    \noalign{\vskip 0.5mm} 
    \hline
    Cuboid & 0.5 & 0.24 & 41  \\
    Pseudo-Halbach & 0.9 & 0.07 & 994 \\
    \hline
    \hline	
  \end{tabular}		
  \label{tab1}
\end{table}

\begin{figure} 	
  \centering
  \includegraphics[width=0.9\linewidth]
  {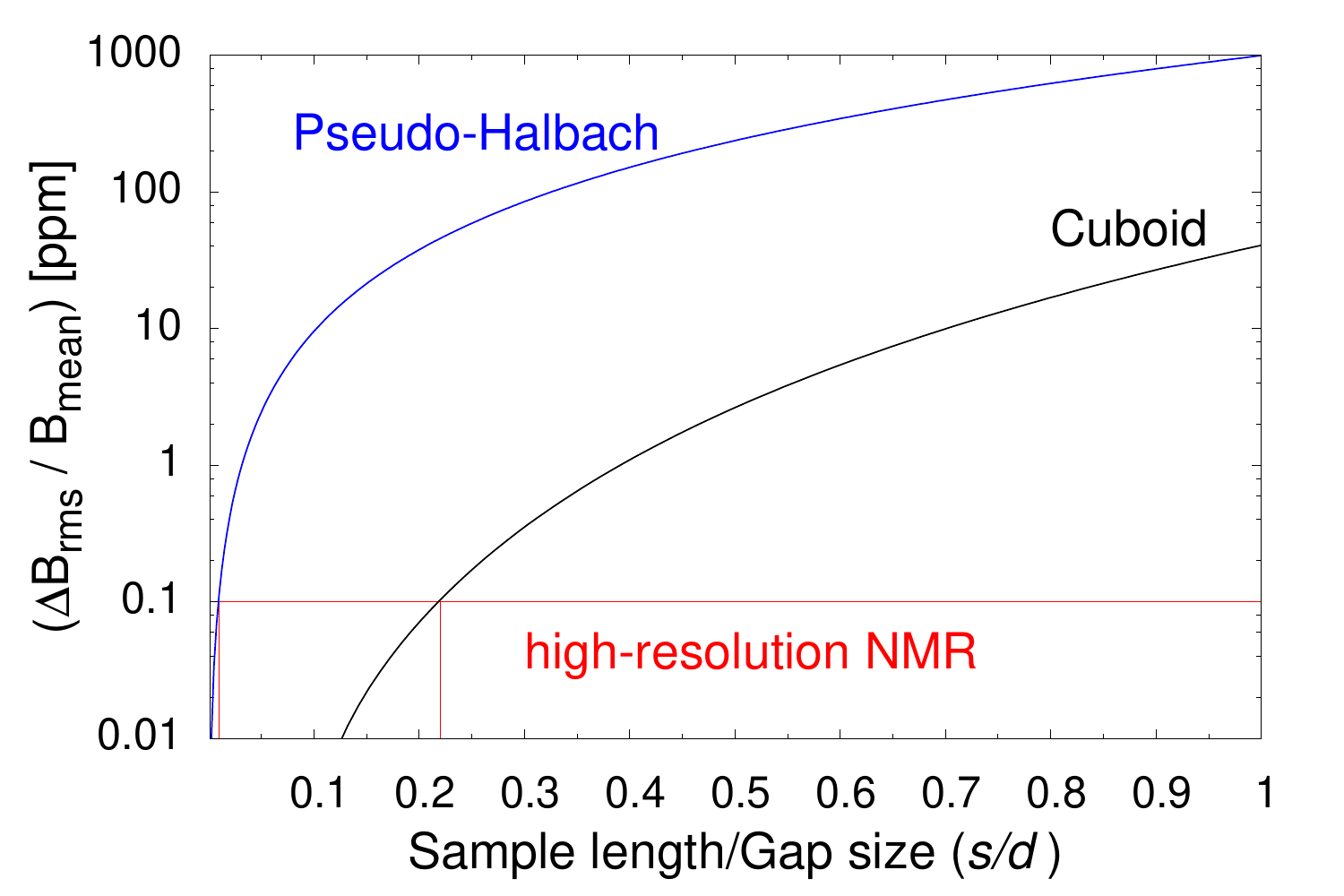}
  \caption{Inhomogeneity of the magnetic field as a function of the
    sample length/gap ratio with a constant configuration width/gap
    ratio of \num{5}. For every sample length, the width of the cuboid
    configurations has been optimised to reach the lowest possible
    inhomogeneity. For a homogeneity reasonable for NMR applications
    of \SI{0.1}{ppm}, the sample length for a cuboid configuration
    needs to be $0.22d$, whereas it has to become unrealistically
    short ($0.01d$) for the pseudo-Halbach configuration.}
  \label{fig6}
\end{figure}

Neither of the two designs reach a field uniformity below
\SI{0.01}{ppm}, which is required for high resolution NMR, so
additional field shimming will remain necessary. However, it is
interesting to analyse whether high resolution NMR systems without
shimming are reachable by reducing the sample length. Therefore, we
optimised the homogeneity of the configuration as a function of sample
lengths, while keeping the outer boundary conditions
intact. Fig.~\ref{fig6} shows how the homogeneity improves with a
reducing ratio of the gap size to the sample length. The cuboid
configuration can indeed reach in theory \SI{0.01}{ppm} with a sample
length of $0.22d$. The pseudo-Halbach configuration however needs an
absurd sample length of $0.01d$ to reach the critical value.

\subsection{Influence of variations in the magnet properties}

To analyse the effect of variation in magnet properties and
positioning on the performance of both designs, we applied a two step
approach. First we analysed the sensitivity of the magnetic field to
the variation in strength and position of a single cubic magnet using
the method of sensitivity matrices. From this we determined that
variations in magnetic moment and angle of magnetisation are most
severe. Focussing on these two parameters only, we analysed the
combined effect of all magnets using a Monte-Carlo
approach. 

\subsubsection{Single cubic magnet}
\begin{figure}
	\centering
	\includegraphics[width=0.2\textwidth]{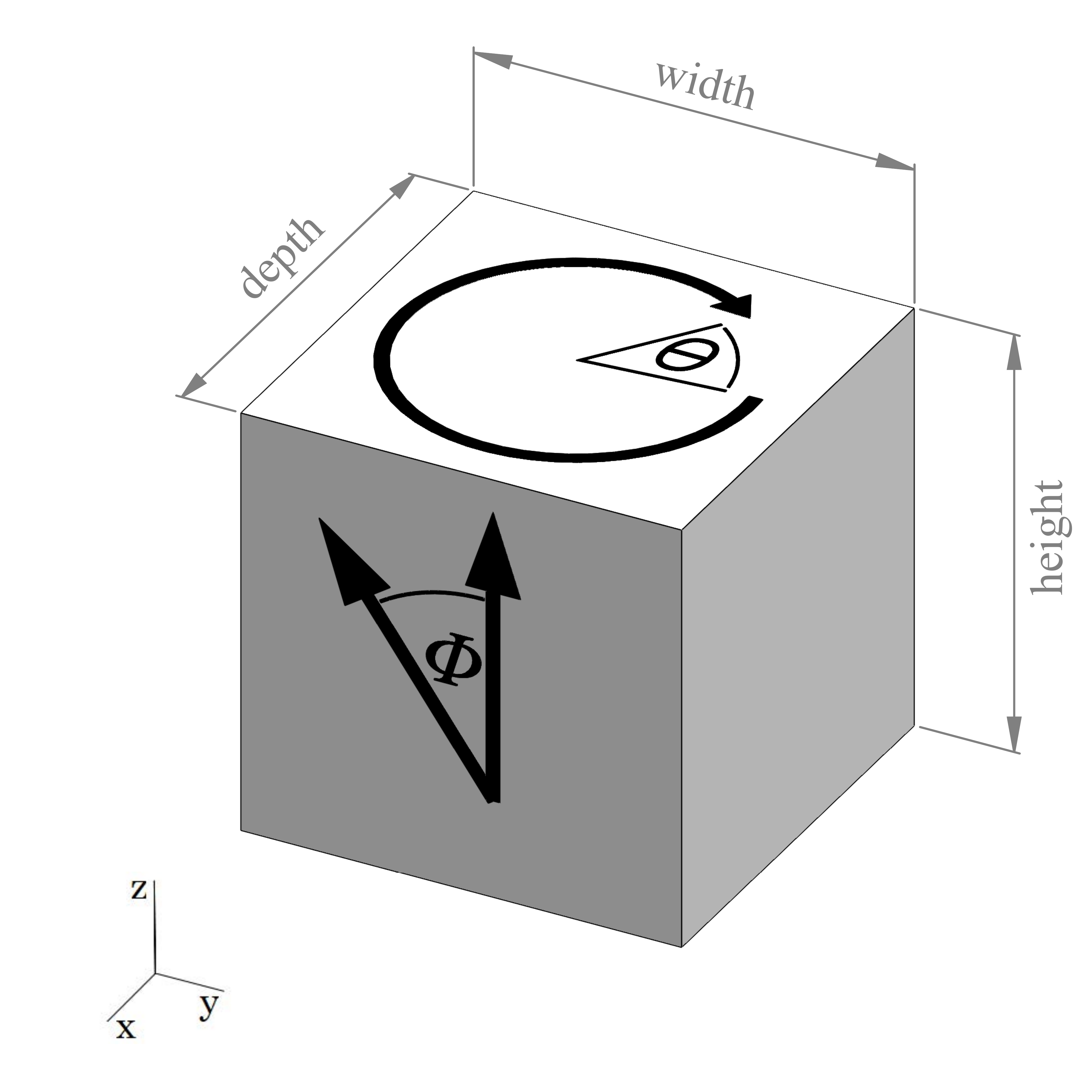}
	\caption{Schematic drawing of the cubic magnet used for
          sensitivity analysis. The length width and depth have a
          nominal size of $a$. The arrows indicate the direction of
          magnetisation. $\phi$ shows the total offset angle, $\theta$
          the offset direction in cylindrical coordinates.}
	\label{fig7}
\end{figure}

We determined the sensitivity of the magnetic stray field of a single
cubic magnet of nominal size $a$ to a variation of the dimensions,
position, and tilt of the magnet, as well as in the magnetisation
strength and angle (Fig.~\ref{fig7}). We consider the field components
$B_{x}, B_{y}, B_{z}$ at a point above the centre of the top (north)
face, at a height of 10\% of the length $a$ of the edge of the magnet.

Table~\ref{tab2reduced} shows the most important elements of the
sensitivity matrix of the magnetic field in the $x$ and $z$-directions
on the $x$-axis at a distance of 0.1$a$, given as percentages of
$B_\text{s}$. Parameters related to the sizes have been varied by
\SI{10}{\percent} of the length of the edge of the cube. Parameters
related to the angle have been varied by \SI{1}{\degree}.
Appendix~\ref{sec:AppendixDesignOptimisation} gives the sensitivities
for $B_{y}$ and the field at $x$=0.1$a$ in Table~\ref{tab2}, as well
as the absolute field values in Figure~\ref{fig8}.

The first row in Table~\ref{tab2reduced} shows that the $B_\text{z}$
component changes proportionally with the magnetisation. Since the
$B_\text{x}$ component is zero (see Figure~\ref{fig8} of
appendix~\ref{sec:AppendixDesignOptimisation} ), variation in
magnetisation has no effect. Similarly, a tilt of the cube or rotation
of the magnetisation around the $y$-axis has a significant influence
only on $B_\text{x}$, but not on $B_\text{z}$. Displacement of the
cube has an effect only on the field components in the direction of
displacement. The effect is relatively small: a \SI{10}{\percent}
variation in position only lead to a \SI{2}{\percent} variation in
field strength.

\begin{table}
  \centering
  \caption{Sensitivity matrix of the magnetic field components ($B_x$
    and $B_z$) at a distance of $0.1a$ above the center of a cubic
    magnet with the edge length $a$ .}
  \begin{tabular}{lcSS}
    \hline 
    \hline
    & Variation & { $B_x$ (\SI{}{\percent}) }&{$B_z$  (\SI{}{\percent})} \\
    \hline
    $M$ & 1\% & 0.00  & \standout{1.00}  \\
    \noalign{\vskip 0.7mm} 
    tilt $x$ & \SI{1}{\degree} & 0.00 & 0.00  \\
    tilt $y$ & \SI{1}{\degree} & \standout{0.61} &  0.00  \\
    \noalign{\vskip 0.7mm} 
    $\phi$ ($\theta$=~\SI{0}{\degree}) & \SI{1}{\degree} & 0.00  & 0.00 \\
    $\phi$ ($\theta$=\SI{90}{\degree}) & \SI{1}{\degree} & \standout{-0.87} & 0.00  \\
    \noalign{\vskip 0.7mm} 
    $x$ & 0.1$a$ & \standout{-1.09} & 0.00 \\
    $z$ & 0.1$a$ & 0.00 & \standout{-2.17}\\
    \hline \hline
  \end{tabular}
  \label{tab2reduced}
\end{table}

\subsubsection{Monte-Carlo simulations}
To analyse the combined effect of all magnets on the field, we
performed a Monte Carlo simulation with \num{50000} draws. Based on
the above analysis of the cube, we consider only variation in the
magnetisation strength and direction. Since for the two configurations
the dimensional variation is smaller than \SI{0.03}{a}, no
dimensional errors were considered. Normal distributions were assumed,
with standard deviations of \SI{1}{\degree} and \SI{1}{\percent} for
strength and angle respectively.

\begin{figure} 
	\centering
	\includegraphics[width=0.9\linewidth]
	{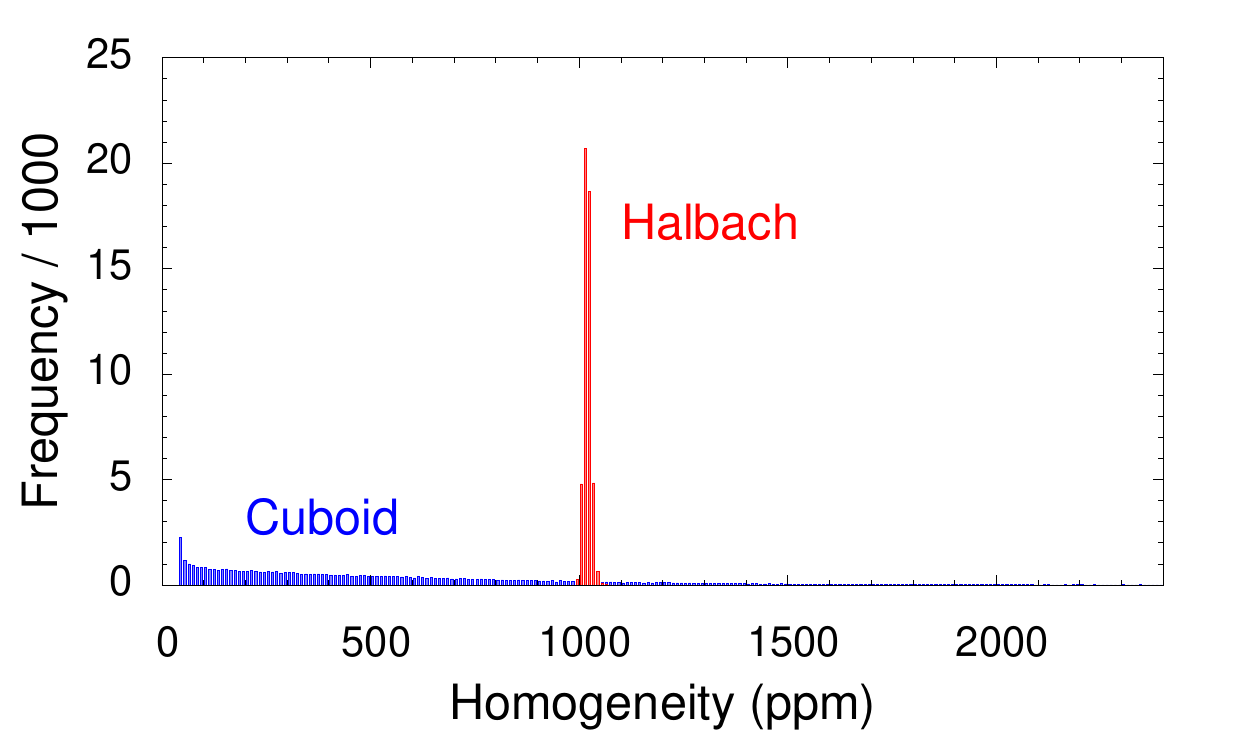}
	\includegraphics[width=0.9\linewidth]
	{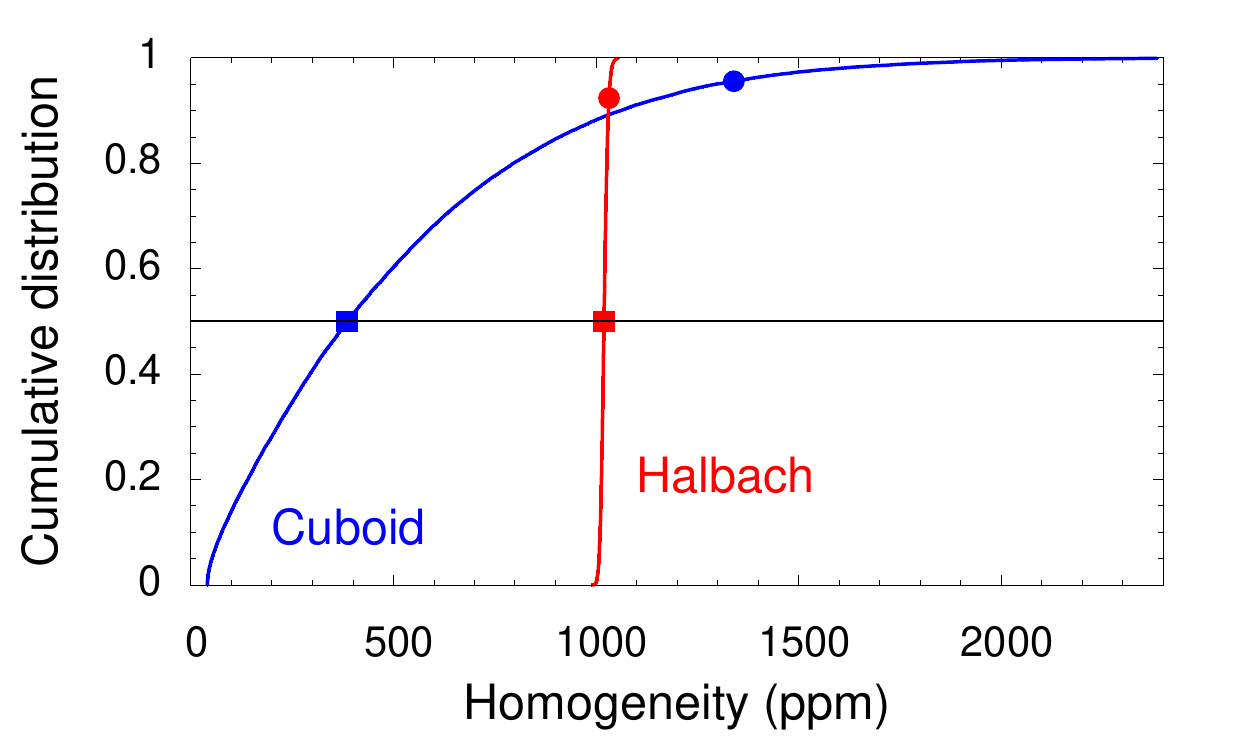}
	\caption{Density plot (top) and cumulative distribution
          function (bottom) of the Monte Carlo simulation for the
          cubic and Pseudo-Halbach configurations. The simulation
          shows that the Pseudo-Halbach configuration has a mean
          homogeneity of \SI{1020}{ppm}, while the cuboid
          configuration reaches \SI{386}{ppm} (Squares in bottom
          figure). The circles indicate the average inhomogeneity
          measured on three implementations of each configuration
          (Table~\ref{tab7}).}
	\label{fig12}
\end{figure}

Fig.~\ref{fig12} shows the distribution and probability plot of the
simulated homogeneities of the magnetic field in the $z$-direction. The
mean homogeneity of the cuboid configuration is \SI{430}{ppm}, the
pseudo-Halbach configuration achieves \SI{1086}{ppm}. However, the
cuboid configuration has a high spread in the homogeneity (standard
deviation \SI{350}{ppm}) while the pseudo-Halbach has a standard
deviation of only \SI{8}{ppm}. With a probability of
\SI{94.4}{\percent}, both the cuboid configuration and the
pseudo-Halbach configuration obtain a homogeneity of \SI{1098}{ppm} or
better.  With a probability of \SI{10}{\percent}, the cuboid
configuration achieves \SI{64}{ppm} whereas the pseudo-Halbach
achieves not less than \SI{1076}{ppm}.

The reason for the strong sensitivity of the cuboid configuration to
magnet variations is largely due to the distribution in magnetisation
direction. Table~\ref{tab:sensitivityCombined}) shows the sensitivity
of the $z$ component of the field at the center and the edge of the
sample to a variation of \SI{1}{\degree} of the magnetisation ($x=d/2$)
angle. At the edge of the sample, the cuboid configuration is ten
times more sensitive to angular variations.

\begin{table}
  \centering
  \caption{Sensitivity to variation in magnetisation angle of $B_z$ at
    the center and edge of the sample for the pseudo-Halbach and
    cuboid configuration}
  \begin{tabular}{lSSl}
    \hline
    \hline
    & {$B_z$ ($x=0$)} & {$B_z$ ($x=d/2$)} & \\ 
    \hline
    Halbach - top/bottom & 0 & 59 &ppm/deg \\ 
    Halbach - side & 0 & 0 & ppm/deg\\ 
    Halbach - corner & 0 & 81 & ppm/deg \\
    Cuboid& 0 & 985 & ppm/deg\\
    \hline
    \hline
  \end{tabular}%
  \label{tab:sensitivityCombined}
\end{table}


\subsection{Verification of simulations with implementations}

Both configurations were assembled and measured three times. The
measurement results are shown in Table~\ref{tab7}. There is a small
spread in the homogeneity of the pseudo-Halbach (mean value of
\SI{1032}{ppm} and standard deviation of \SI{90}{ppm}). A larger
variation was found for the cuboid configuration (\SI{1340}{} and
\SI{800}{ppm}). (Raw data of the six systems is listed in
Table~\ref{tab6} of appendix~\ref{sec:AppendixMeasurements} ).

The three implementations represent a draw from the \SI{50000} Monte
Carlo simulations shown in figure Fig.~\ref{fig12}. From these curves
we can estimate that the chance to realise a cuboid configuration with
an inhomogeneity \emph{as bad as} \SI{1340}{ppm} is in the order of
\SI{5}{\percent} (blue dot in figure). Similarly, a pseudo-Halbach
configuration with a inhomogeneity of \SI{1032}{ppm} or worse has a
chance of \SI{8}{\percent} of occurring. These likelihoods are low, but
not unrealistic. More implementations would be required to determine
whether other variation than magnetisation strength and direction
should be considered.

In general, the pseudo-Halbach configuration has a more predictable
field profile, which makes this design more favourable for industrial
applications than the cuboid configuration. Since shimming is needed
anyway, a measurement of the field profile is not necessary. We
therefore recommend restricting the use of the cuboid configurations
to research systems, where selecting the magnets and measuring the
final assembly is feasible.

\begin{table}
  \centering
  \caption{Mean (standard deviation) of the homogeneity of the field
    in the $z$-direction of measured and simulated magnet
    configurations}
  \begin{tabular}{lSS}
    \hline
    \hline
    & {Measured } & {Simulated }\\
    & {Inhomogeneity} & {Inhomogeneity}\\
    &  {$[$ppm$]$} & {$[$ppm$]$} \\
    \noalign{\vskip 0.5mm} 
    \hline
    Cuboid & 1340(800) & 386\\ 
    Pseudo-Halbach & 1032(90) & 1020 \\
    \hline
    \hline
  \end{tabular}
  \label{tab7}
\end{table}



\section{Conclusion}
We have investigated the effect on the homogeneity of the field of permanent magnet configurations for mobile NMR applications of variations in the properties of the magnets.
We measured the variations in the magnetisation and magnetisation angle of permanent magnets but could not observe a decisive difference between the manufacturers, materials, or magnet coatings. On average, the standard deviation of the magnetisation is less than \SI{1}{\percent} and for the variations in the magnetisation angle it is less than \SI{1}{\degree}. 

We compared a cuboid and a pseudo-Halbach magnet configuration, in terms
of their field strength and field homogeneity, for our optimised boundary
conditions, in which the sample length $s$ is equal to the gap size $d$ and the
whole configuration should fit in a box with an edge length five times
the gap size. For a fixed parameter set, assuming perfectly magnetised
magnets, the field in the centre of the cuboid configuration is
\num{0.5}$B_\text{s}$ and its homogeneity is \SI{41}{ppm}. For the same
boundary conditions, the pseudo-Halbach configuration achieves a higher
field (\num{0.9}$B_\text{s}$) in the centre but less
homogeneity (\SI{994}{ppm}). It is worth mentioning that the
pseudo-Halbach configuration has a much lower stray field, and so less interference
with the environment, than the cuboid configuration.

For samples with a length the same as the gap size, the theoretical homogeneity of both configurations is above the sub-ppm range, which is necessary to produce a high resolution spectrum. Optimising the homogeneity for shorter samples while respecting the maximum outer dimensions yields in a much better homogeneity. Using a sample length of $0.22d$ improves the homogeneity from \SI{41}{} to \SI{0.1}{ppm} for the cuboid configuration, whereas the pseudo-Halbach configuration would need a impractical sample length of $0.01d$. 

We analysed the effect of the variation in magnetic properties on the uniformity of the generated fields.
The sensitivity matrix shows that the magnetisation, magnetisation angle, and tilt have the most significant influence on the magnetic field.
Positioning errors mainly change the field, in case the positioning variation is in the same direction as the field.
Theoretically, the cuboid has good homogeneity (on average \SI{430}{ppm}), but the effect of variation in the magnets'  properties  is large (standard deviation \SI{350}{ppm}).
The pseudo-Halbach configuration has worse homogeneity
(\SI{1080}{ppm}), but is \num{44} times less sensitive to variation in
the properties of the magnet.

We verified the modelled field inhomogeneities with three realisations
for each of the two designs. The average inhomogeneity agree within
measurement error with the model.

Based on our analysis, we advise using the cuboid configuration for
scientific use, where it is possible to preselect the permanent
magnets and the external stray field is not a big issue.  Mechanical
shimming of this configuration can be done, changing the distance between
the magnets (counteracting magnetisation differences) or by tilting
the magnet (counteracting magnetisation angle variations).  Using
rather large magnets helps to achieve the homogeneity needed for NMR
measurements.  If preselecting the magnets is not an option, we
recommend the pseudo-Halbach configuration, which has a more robust
homogeneity regarding variations in the magnetisation and angle.  The
field profile of this configuration is predictable, which makes it
easier to shim afterwards to achieve the field homogeneity needed for
NMR applications.  Also the lower stray field makes this configuration
easier to handle and therefore more favourable especially for
industrial applications.


\section*{Acknowledgements}

This work is part of the research programme FLOW+ with project number 15025, which is (partly) financed by the Dutch Research Council (NWO). The authors thank Jankees Hogendoorn and Lucas Cerioni of Krohne New Technologies BV for their input and support.

\bibliographystyle{IEEEtran}
\bibliography{../../paperbase/paperbase}
\clearpage
\section{Appendix}

\appendices

\section{Distribution of permanent magnet properties}
\label{sec:appendixMagnetDistribution}

We measured the magnetisation strength and orientation for a range of
commercially available permanent magnets (listed in table~\ref{tab10}
of main text). The results are summarized in table~\ref{tab3} in the main
text. The underlying data is reported in this appendix.

\begin{figure}
  \centering	\includegraphics[width=1\linewidth]{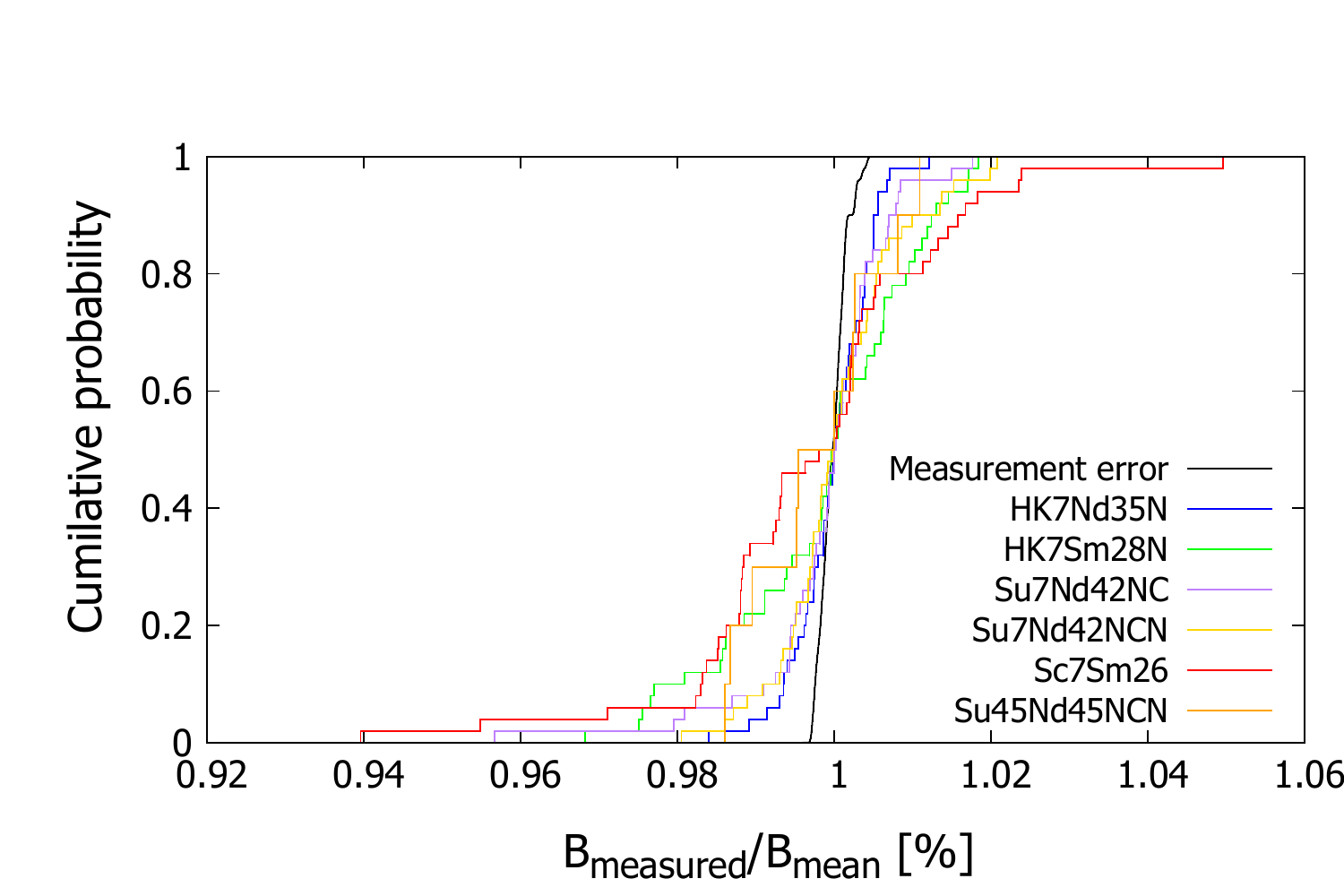}
  \caption{Measured cumulative distribution of the residual flux
    density for a range of commercial magnets. The black line
    indicates our sensitivity limit. The sensitivity limit has
    been obtained from measuring \num{50} times the same magnet,
    indicated by the black line. We measured \num{10} different
    magnets with a diameter of $d$=\SI{45}{mm} and a height of
    $h$=\SI{30}{mm} (orange), and \num{50} magnets with a size
    of \SI{7x7x7}{mm} for each of the other kinds of material or
    manufacturer. On average, commercial magnets have a
    magnetization variation of less than \SI{1}{\percent}.}
  \label{fig10}
\end{figure}

Figure~\ref{fig10} shows the cumulative distribution of
the residual flux density, normalized to the mean value. The
measurement uncertainty is estimated from the cumulative distribution
for \num{50} measurements of the same magnet (black curve). The
distribution in magnetisation of the commercial magnets exceeds our
measurement uncertainty. The standard deviation is in the order of
\SI{1}{\percent}.

\begin{figure}
  \centering
  \includegraphics[width=1\linewidth]{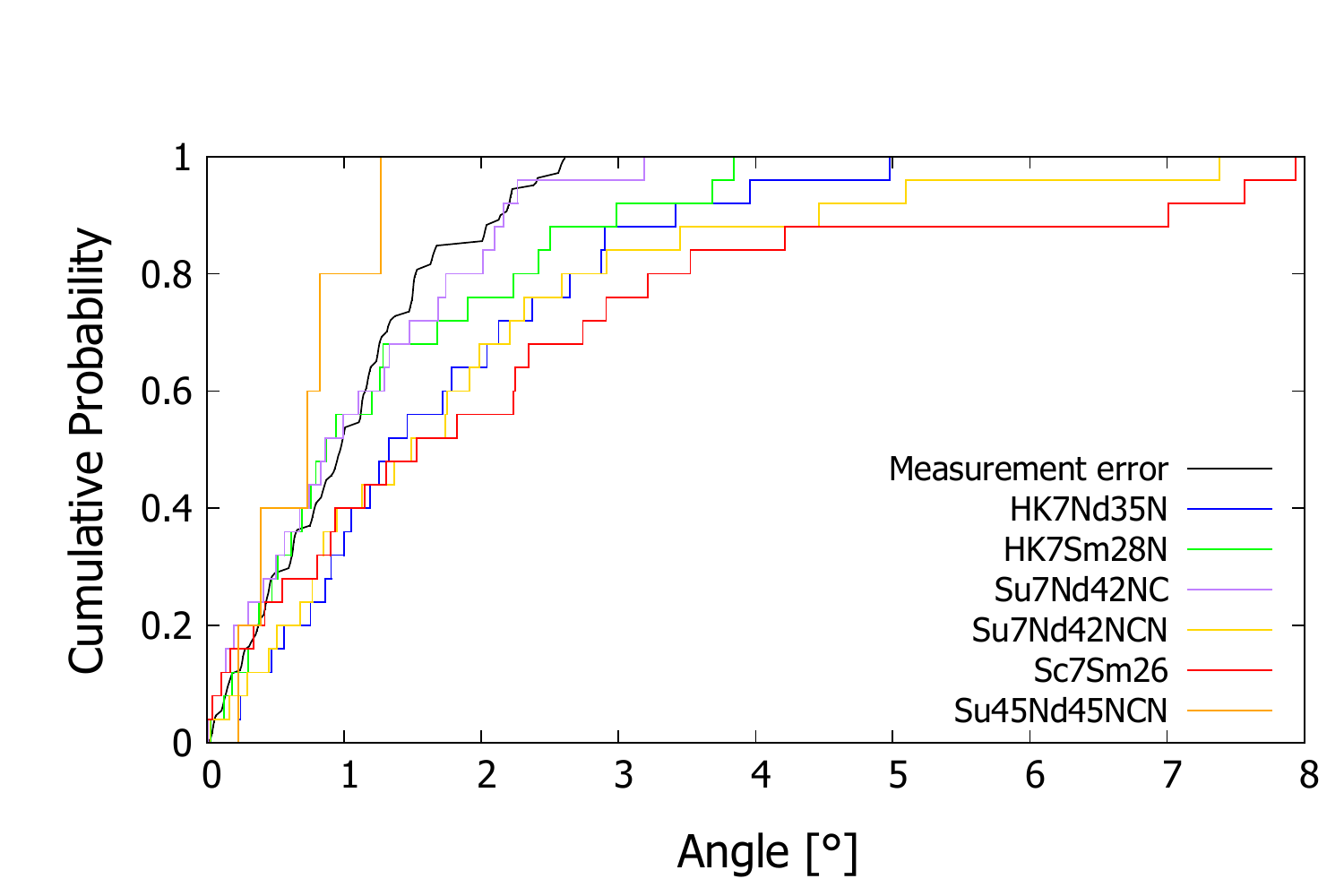}
  \caption{Measured cumulative distribution of the field direction
    with respect to the $z$-axis ($\phi$ in figure~\ref{fig7}) for a
    range of commercial magnets. On average, commercial magnets have a
    field direction variation of less than \ang{1}.}
  \label{fig11}
\end{figure}

\begin{figure}
	\centering	\includegraphics[width=0.9\columnwidth]{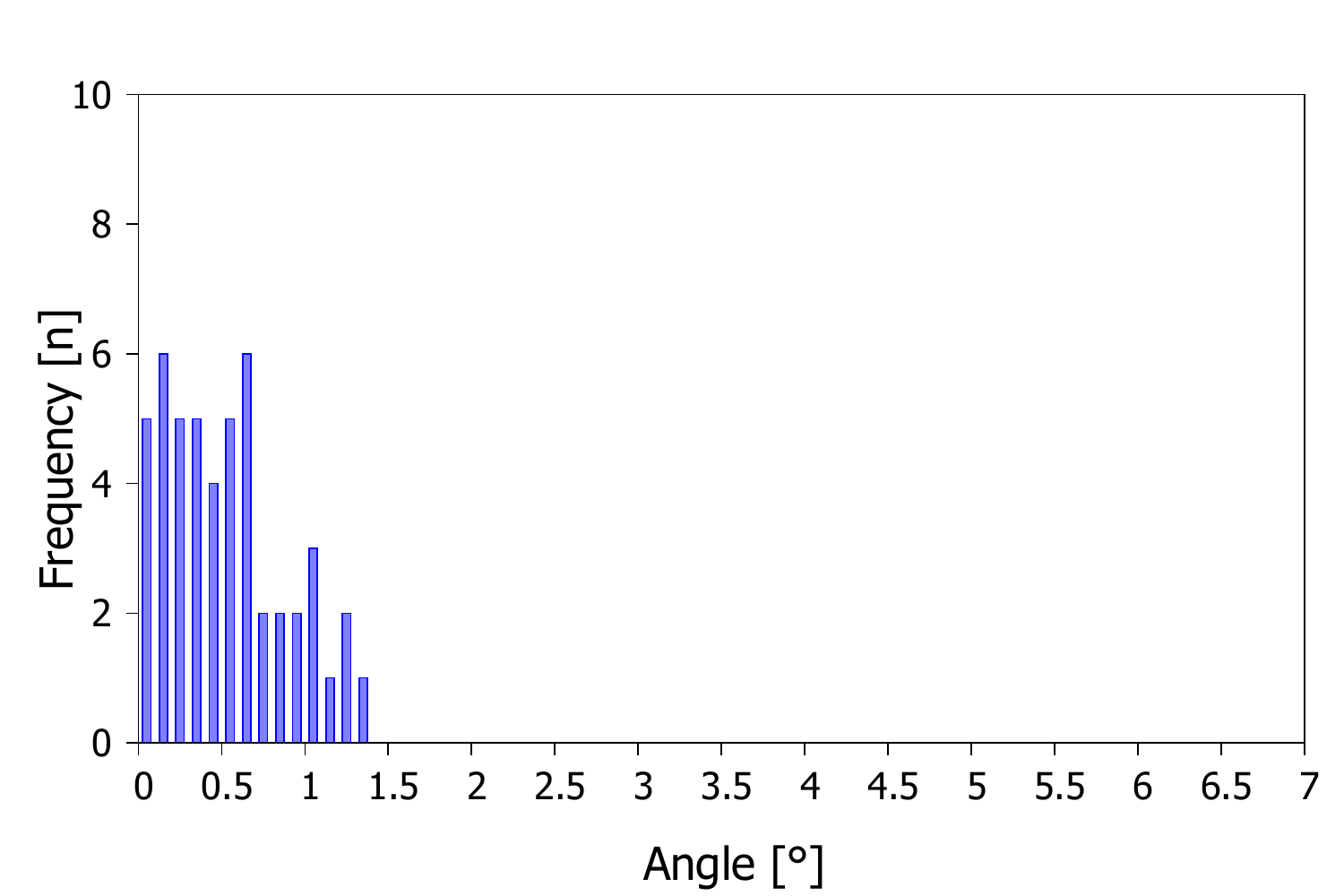}
	\caption{The offset angle from the same magnet has been
		measured \num{50} times resulting in a standard deviation of
		\SI{0.645}{\degree}.\label{fig:mean50deg}}
\end{figure}

In a similar fashion, the variation in field direction was measured
(Figure~\ref{fig11}). The base uncertainly measurement is again shown
as a black curve. The histogram presentation is shown in
figure~\ref{fig:mean50deg}. The HKCM magnets appear to have a
smaller angular variation that the other small magnets. The angular
varation of type HK7Nd35N cannot be measured accurately by our
method. The variation of the angular variation of the big Supermagnete
magnet (Su45Nd45NCN) was assesed only on \num{10} magnets. It appears
however that the variation is well below our measurement uncertainty.

Figure~\ref{fig:mean50deg} shows the offset angle from the
same magnet, which has been measured \num{50} times resulting in a
standard deviation of
\SI{0.645}{\degree}.

Table~\ref{tab:AngleVariations} and~\ref{tab:MagVariations} summarize the
measured angle and magnetization variations with confidence intervals.

\begin{figure*}[t]
	\centering
	\includegraphics[width=0.6\textwidth]{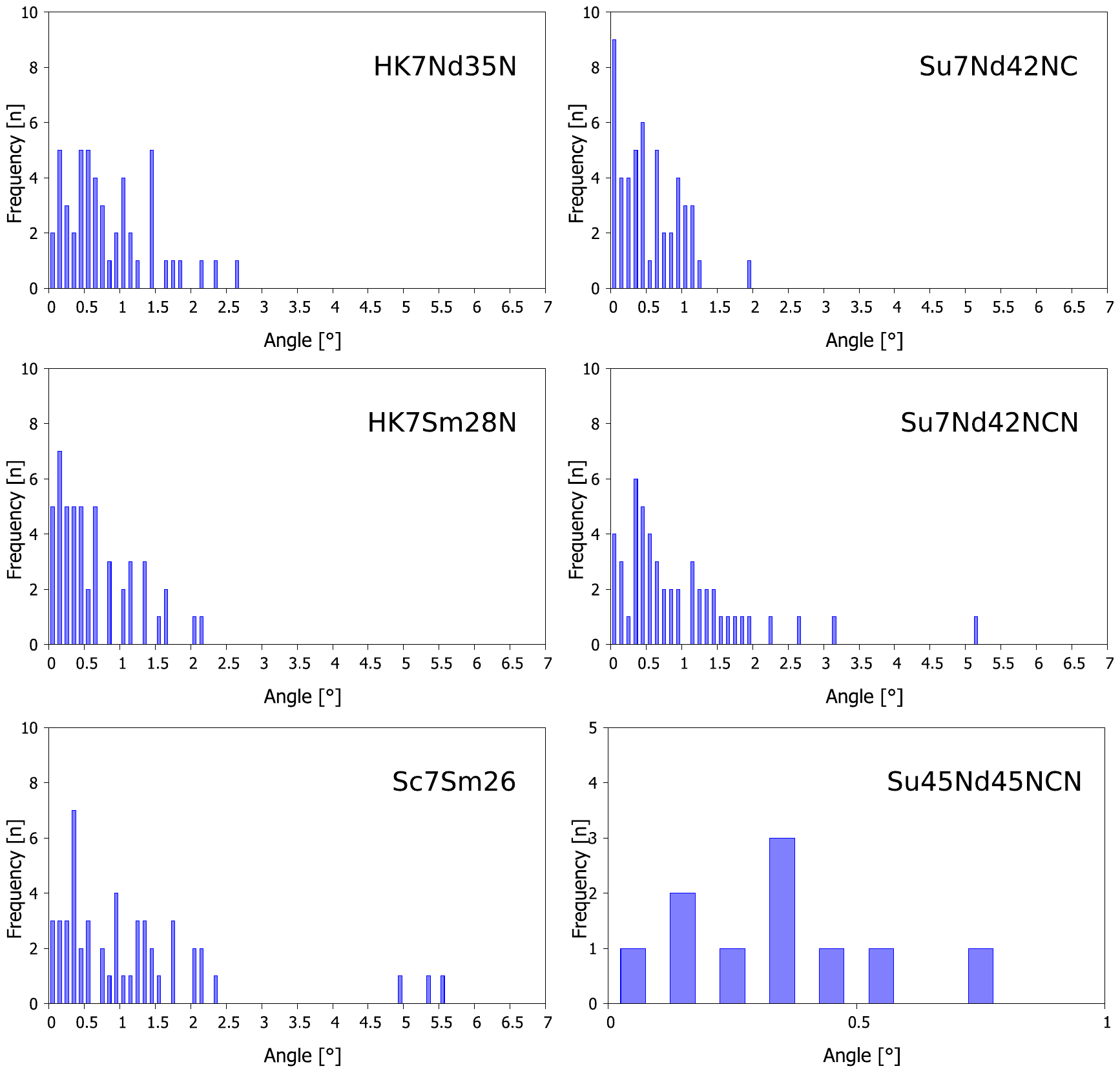}
	\caption{Measured field angles for the range of commercial magnets
		investigated }
	\label{fig:AngleSummary}
\end{figure*}

Figures~\ref{fig:AngleSummary} shows the raw measured angular
variations for the series of commercial magnets investigated.

\newpage

\clearpage

\begin{table*}[b]
	\caption{Measured angle variations.\label{tab:AngleVariations}}
	\centering
	\begin{tabularx}{0.8\textwidth}{p{3cm}p{3cm}p{3cm}SSS}
		\hline
		\hline
		&&&& \multicolumn{2}{c}{confidence interval}\\
		\cmidrule(lr){5-6}
		& Material & Coating & {std $\rho$ [\si{\degree}]} & \SI{95}{\percent}
		& \SI{68.27}{\percent} \\
		\hline
		Measurement error & NdFeB N45 & Ni-Cu-Ni & 0.65 & 0.11 &  0.06  \\ 
		Supermagnete & NdFeB N42 & Ni-Cu-Ni & 1.33 & 0.2 & 0.11  \\
		Supermagnete & NdFeB N42 & Ni-Cu & 0.69 & 0.113 & 0.06  \\ 
		HKCM & NdFeB N35 & Ni & 1.05 & 0.17 & 0.09  \\ 
		HKCM & Sm2Co17 YXG28 & Ni & 0.85 & 0.14 & 0.07  \\ 
		Schallenkammer Magnetsysteme& Sm2Co17 YXG-26H & - & 1.72 & 0.28 & 0.14  \\ 
		Supermagnete (Cylinder) & NdFeB N45 & Ni-Cu-Ni & 0.42 &  0.13 &  0.07 \\
		\hline
		\hline
	\end{tabularx} 
\end{table*}

\begin{table*}[b]
	\caption{Measured magnetization variations.\label{tab:MagVariations}}
	\centering
	\begin{tabularx}{0.8\textwidth}{p{3cm}p{3cm}p{2.5cm}SSS}
		\hline
		\hline
		& Material & Coating & {B [T}] & {$\sigma$} &
		{\SI{68.27}{\percent}}  \\
		& & & & & {confidence}\\
		\hline
		Measurement error & NdFeB N42 & Ni-Cu-Ni & 0.5425 & 0.1709 & 0.0002 \\ 
		Supermagnete & NdFeB N42 & Ni-Cu-Ni & 0.4196 & 0.8631 & 0.1374 \\ 
		Supermagnete & NdFeB N42 & Ni-Cu & 0.5187 & 0.7965 & 0.0669 \\ 
		HKCM & NdFeB N35 & Ni & 0.4503 & 0.9208 & 0.0773 \\ 
		HKCM & Sm2Co17 YXG28 & Ni & 0.4362 & 0.5025 & 0.0422 \\ 
		Schallenkammer Magnetsysteme & Sm2Co17 YXG-26H & - & 0.3852 & 1.2024 & 0.1010 \\ 
		Supermagnete (Cylinder) & NdFeB N45 & Ni-Cu-Ni & 0.3687 & 1.7225 & 0.1447 \\
		\hline
		\hline
	\end{tabularx}
\end{table*}










\clearpage
\section{Design optimisation}
\label{sec:AppendixDesignOptimisation}
\begin{figure}[htb]
	\centering
	\includegraphics[width=1\linewidth]{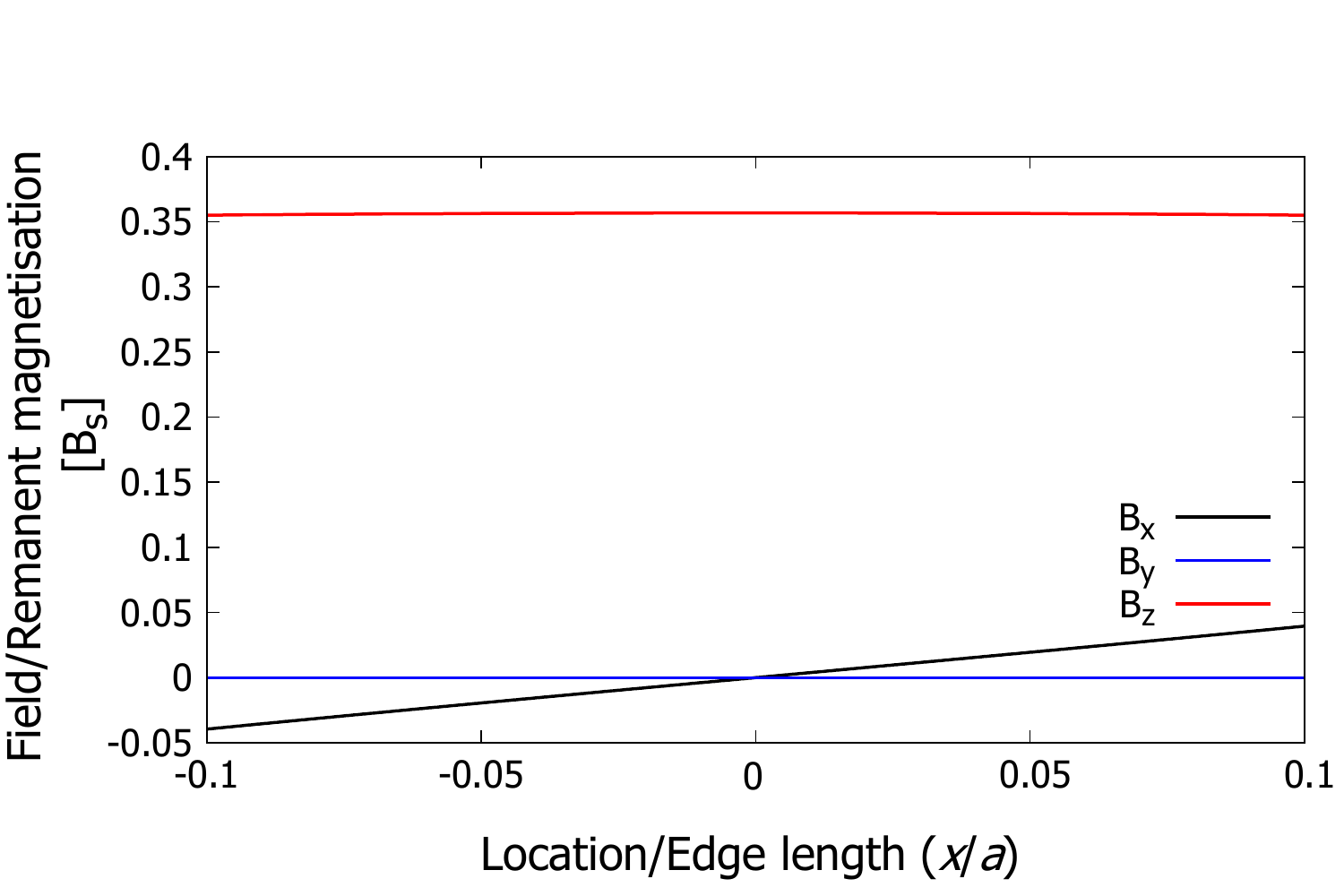}
	\caption{Magnetic field ($x,y,z$) above a cuboid magnet with
		the edge length $a$ along the $x$-axis at a distance of
		\num{0.1}$a$.}
	\label{fig8}
\end{figure}

Fig.~\ref{fig8} shows that for a cubic magnet, magnetized in
the $z$-direction, there is no field in the $y$-direction ($B_{y}$) along the
$x$-axis at a distance of \num{0.1}$a$ from the surface. The field
$B_{x}$ is zero in the centre of the magnet and rises linearly with a
slope of \num{0.04}$B_\text{s}/a$ in the positive
$x$-direction, with $B_\text{s}$ the saturation magnetization of the
magnet material [T]. The field $B_{z}$ is \num{0.357}$B_\text{s}$ in the centre which drops to \num{0.355}$B_\text{s}$ at $x=0.1a$.

%
%
%


In ~\cref{tab:sensBx,tab:sensBy,tab:sensBz} the sensitivity matrix of $B_x, B_y, B_z$ above a cuboid magnet with the edge length $d$ along $x$-axis at a distance of $0.1d$ is shown.\\

\begin{table*}
	\centering
	\caption{Sensitivity matrix of the magnetic field
		($x$,$y$,$z$), given in the change of
		$B_\text{s}$ at the same position in [\SI{}{\percent}] above a cuboid magnet
		with the edge length $a$ along the $x$-axis at a distance of
		$0.1a$. Variations in the magnetization angle
		and tilting the magnet perpendicular to the simulated
		axis affect the magnetic field significantly. Placement errors have
		an influence if parallel to the field. Variations in the dimensions have a
		minor effect.}
	\begin{tabularx}{0.8\textwidth}{ll@{\extracolsep{\fill}}SSSSSS}
		\hline \hline
		& & \multicolumn{2}{c}{$B_{x}$} & \multicolumn{2}{c}{$B_{y}$} &
		\multicolumn{2}{c}{$B_{z}$} \\
		\cmidrule(lr){3-4} \cmidrule(lr){5-6}  \cmidrule(lr){7-8} 
		& Variation & {$x=0$} & {$x=0.1a$} &{$x=0$} & {$x=0.1a$} &{$x=0$} & {$x=0.1a$} \\
		\hline
		$M$ & 1\% & 0.00 & \standout{1.10} & 0.00 & 0.00 & \standout{1.00} & \standout{0.99} \\
		\noalign{\vskip 0.7mm} 
		tilt $x$ & \SI{1}{\degree} & 0.00 & 0.00 & \standout{-0.61} & \standout{-0.62} & 0.00 & 0.00 \\
		tilt $y$ & \SI{1}{\degree} & \standout{0.61} & \standout{0.51} & 0.00 & 0.00 & 0.00 & \standout{-0.48} \\
		\noalign{\vskip 0.7mm} 
		$\phi$ ($\theta$=~\SI{0}{\degree}) & \SI{1}{\degree} & 0.00 & 0.00 & 0.00 & 0.00 & 0.00 & 0.00 \\
		$\phi$ ($\theta$=\SI{90}{\degree}) & \SI{1}{\degree} & \standout{-0.87} & \standout{-0.88} & \standout{-0.87} & \standout{-0.86} & 0.00 & 0.19 \\
		\noalign{\vskip 0.7mm} 
		$x$ & 0.1$a$ & \standout{-1.09} & \standout{-1.15} & 0.00 & 0.00 & 0.00 & 0.10 \\
		$y$ & 0.1$a$ & 0.00 & 0.00 & \standout{1.09} & \standout{1.07} & 0.00 & 0.00\\
		$z$ & 0.1$a$ & 0.00 & -0.10 & 0.00 & 0.00 & \standout{-2.17} &\standout{-2.22} \\
		\noalign{\vskip 0.7mm} 
		height & 0.1$a$ & 0.00 & 0.02 & 0.00 & 0.00 & \standout{0.23} & \standout{0.23} \\
		depth & 0.1$a$ & 0.00 & \standout{-0.18} & 0.00 & 0.00 & -0.01 & 0.01 \\
		width & 0.1$a$ & 0.00 & 0.05 & 0.00 & 0.00 & -0.01 & -0.01 \\
		\hline \hline
	\end{tabularx}
	\label{tab2}
\end{table*}

\label{sec:sensitivityCuboid}
An indication of why the cuboid configuration has a much higher standard
deviation than the pseudo-Halbach configuration can be seen
from the sensitivity matrices of the $z$-field. We chose to show how
the field in the centre and at $x$=$d/2$ changes for a magnetization
difference of \SI{1}{\percent} and an offset magnetization direction
of \SI{1}{\degree} each in the direction which creates the highest
field difference at both locations. The Halbach configuration consists
of 8 magnets: 4 corner magnets, 2 at the side, and 1 each on top and
bottom. Adding up the sensitivity values of all the magnets results in a
difference of \SI{314}{ppm} between the $z$-field at $x$=0 and
$x$=$d/2$. The cuboid shows a significantly higher difference of
\SI{1970}{ppm}.

%

\begin{table*}[h!]
	\centering
	\caption{Sensitivity matrix of $B_x$ above a cuboid magnet with the
		edge length $d$ along $x$-axis at a distance of $0.1d$}
	\label{tab:sensBx}
	\begin{tabularx}{0.8\textwidth}{l*{9}{S[table-format=4.2]}}
		\hline
		\hline
		{$B_x(x)$}  & {$-0.1d$} & {$-0.075d$} &
		{$-0.05d$} & {$-0.025d$} & 0 & {$0.025d$} &
		{$0.05d$} & {$0.075d$} & {$0.1d$} \\
		{[\num{e-6}$B_\text{s}$]} & \\
		\hline
		$B_x$ & -39520.0 & -29365.0 & -19448.0 & -9685.3 & 0 & 9685.3 & 19448.0 & 29365.0 & 39520.0 \\ 
		$\Delta M$  & -394.6 & -294.1 & -194.8 & -96.9 & 0 & 96.9 & 194.8  & 294.1 & 394.6 \\ 
		$\Delta$tilt $x$ & 0 & 0 & 0 & 0 & 0 & 0 & 0 & 0 & 0 \\ 
		$\Delta$tilt $y$ & 1822.1 & 1972.9 & 2086.0 & 2148.8 & 2174.0 &
		2148.8 & 2086.0 & 1972.9 & 1822.1 \\
		$\Delta \phi$=\SI{1}{\degree}, $\theta$=\SI{0}{\degree} & 0 & 0 & 0 & 0 & 0 & 0 & 0 & 0 & 0 \\ 
          $\Delta \phi$=\SI{1}{\degree}, $\theta$=\SI{90}{\degree} & -3129.0 & -3116.5 & -3116.5 & -3116.5 & -3116.5 & -3116.5 & -3116.5 & -3116.5 & -3129.0 \\ 
		$\Delta x$ & -4121.8 & -4008.7 & -3933.3 & -3883.0 & -3870.4 & -3883.0 & -3933.3 & -4008.7 & -4121.8 \\ 
		$\Delta y$ & 0 & 0 & 0 & 0 & 0 & 0 & 0 & 0 & 0 \\ 
		$\Delta z$ & 360.7 & 257.6 & 165.9 & 81.6 & 0 & -81.6 & -165.9 & -257.6 & -360.6 \\ 
		$\Delta$ height & -88.7 & -66.8 & -44.7 & -22.4 & 0 & 22.4 & 44.7 & 66.8 & 88.7 \\ 
		$\Delta$ depth & 647.2 & 478.8 & 315.4 & 157.1 & 0 & -157.1 & -315.4 & -478.8 & -647.2 \\ 
		$\Delta$ width & -182.2 & -137.0 & -91.4 & -45.7 & 0 & 45.7 & 91.4 & 137.0 & 182.2 \\
		\hline
		\hline
	\end{tabularx}
\end{table*}

\begin{table*}[h!]
	\caption{Sensitivity matrix of$B_y$ above a cuboid magnet with the
		edge length $d$ along $x$-axis at a distance of $0.1d$}
	\label{tab:sensBy}
	\centering
	\begin{tabularx}{0.8\textwidth}{l*{9}{S[table-format=4.2]}}  
		\hline
		\hline
		{$B_y(x)$}  & {$-0.1d$} & {$-0.075d$} &
		{$-0.05d$} & {$-0.025d$} & 0 & {$0.025d$} &
		{$0.05d$} & {$0.075d$} & {$0.1d$} \\
		{[\num{e-6}$B_\text{s}$]} & \\
		\hline
		$B_y$ & 0 & 0 & 0 & 0 & 0 & 0 & 0 & 0 & 0  \\
		$M$ & 0 & 0 & 0 & 0 & 0 & 0 & 0 & 0 & 0 \\ 
		tilt $x$ & -2211.7 & -2199.1 & -2186.5 & -2174.0 & -2174.0 & -2174.0 & -2186.5 & -2199.1 & -2211.7 \\ 
		tilt $y$ & 0 & 0 & 0 & 0 & 0 & 0 & 0 & 0 & 0 \\ 
		$\Delta \phi$=\SI{1}{\degree}, $\theta$=\SI{0}{\degree} & -3066.2 & -3091.3 & -3103.9 & -3103.9 & -3116.5 & -3103.9 & -3103.9 & -3091.3 & -3066.2 \\ 
		$\Delta \phi$=\SI{1}{\degree}, $\theta$=\SI{90}{\degree}  & 0 & 0 & 0 & 0 & 0 & 0 & 0 & 0 & 0 \\ 
		$\Delta x$ & 0 & 0 & 0 & 0 & 0 & 0 & 0 & 0 & 0 \\ 
		$\Delta y$ & 3807.6 & 3832.7 & 3857.9 & 3870.4 & 3870.4 & 3870.4 & 3857.9 & 3832.7 & 3807.6 \\ 
		$\Delta z$ & 0 & 0 & 0 & 0 & 0 & 0 & 0 & 0 & 0 \\ 
		$\Delta$ height & 0 & 0 & 0 & 0 & 0 & 0 & 0 & 0 & 0 \\ 
		$\Delta$ depth & 0 & 0 & 0 & 0 & 0 & 0 & 0 & 0 & 0 \\ 
		$\Delta$ width & 0 & 0 & 0 & 0 & 0 & 0 & 0 & 0 & 0 \\
		\hline
		\hline
	\end{tabularx}
\end{table*}

\begin{table*}[h!]
	\caption{Sensitivity matrix of $B_z$ above a cuboid magnet with the
		edge length $d$ along $x$-axis at a distance of 0.1\textit{d} }
	\label{tab:sensBz}
	\centering
	\begin{tabularx}{0.8\textwidth}{l*{9}{S[table-format=4.2]}}  
		\hline
		\hline
		{$B_z(x)$}  & {$-0.1d$} & {$-0.075d$} &
		{$-0.05d$} & {$-0.025d$} & 0 & {$0.025d$} &
		{$0.05d$} & {$0.075d$} & {$0.1d$} \\
		{[\num{e-6}$B_\text{s}$]} & \\
		\hline
		$B_z$  & 354950 &  355720 & 356250 & 356550 & 356660 & 356550 & 356250 & 355720 & 354950 \\ 
		$M$ & 3543.7 & 3556.3 & 3556.3 & 3568.8 & 3568.8 & 3568.8 & 3556.3 & 3556.3 & 3543.7 \\ 
		tilt $x$ & 0 & 0 & 0 & 0 & 0 & 0 & 0 & 0 & 0 \\ 
		tilt $y$ & 1696.5 & 1269.2 & 844.5 & 422.2 & 0 & -422.2 & -844.5 & -1269.2 & -1696.5 \\ 
		$\Delta \phi$=\SI{1}{\degree}, $\theta$=\SI{0}{\degree} & 0 & 0 & 0 & 0 & 0 & 0 & 0 & 0 & 0 \\ 
		$\Delta \phi$=\SI{1}{\degree}, $\theta$=\SI{90}{\degree} & -689.9 & -512.7 & -339.3 & -169.6 & 0 & 169.6 & 339.3 & 512.7 & 689.9 \\ 
		$\Delta x$ & -360.7 & -257.6 & -165.9 & -81.6 & 0 & 81.6 & 165.9 & 257.6 & 360.7 \\ 
		$\Delta y$ & 0 & 0 & 0 & 0 & 0 & 0 & 0 & 0 & 0 \\ 
		$\Delta z$ & -7929.4 & -7841.4 & -7778.6 & -7753.5 & -7740.9 & -7753.5 & -7778.6 & -7841.4 & -7929.4 \\ 
		$\Delta$ height & 821.8 & 826.9 & 830.6 & 833.2 & 833.2 & 833.2 & 830.6 & 826.9 & 821.8 \\ 
		$\Delta$ depth & 40.0 & 8.4 & -13.2 & -25.8 & -30.0 & -25.8 & -13.2 & 8.4 & 40.0 \\ 
		$\Delta$ width & -33.0 & -31.7 & -30.7 & -30.2 & -30.0 & -30.2 & -30.7 & -31.7 & -33.0 \\
		\hline
		\hline
	\end{tabularx}
\end{table*}


\clearpage
\section{Measurements}
\label{sec:AppendixMeasurements}

\begin{figure}[htb] 
	\centering
	\includegraphics[width=0.9\columnwidth]{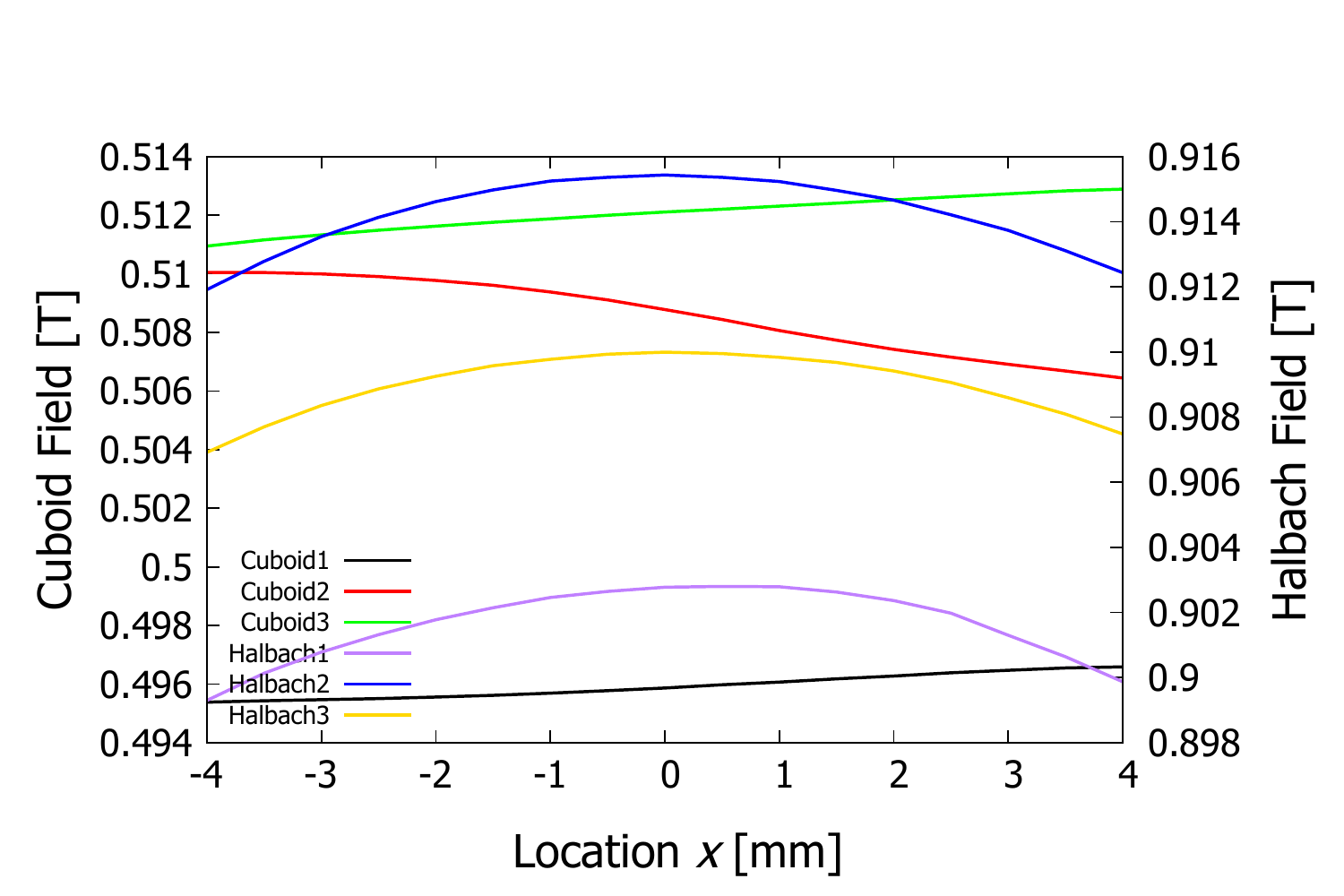}
	\caption{Measured magnetic field ($z$) of Cuboid and Pseudo-Halbach configuration along $x$-axis.\label{fig:measurementally2V2}
	}
\end{figure}

Figure~\ref{fig:measurementally2V2} shows the measured magnetic field ($B_z$) of Cuboid and Pseudo-Halbach configuration along the $x$-axis for $d$ = \SI{8}{mm}.
Table~\ref{tab6} shows the homogeneities of the measured configurations.

\begin{table}[h]
  \centering
  \caption{Measured homogeneity of cuboid and pseudo-Halbach configurations}
  \begin{tabular}{lS}
    \hline
    \hline
    & {Inhomogeneity}\\
    & $[$ppm$]$ \\
    \noalign{\vskip 0.5mm} 
    \hline
    Cuboid 1 & 748(3) \\ 
    Cuboid 2 & 2250(3) \\ 
    Cuboid 3 & 1021(3) \\ 
    Pseudo-Halbach 1 & 1088(3) \\ 
    Pseudo-Halbach 2 & 1081(3) \\ 
    Pseudo-Halbach 3 & 929(3) \\
    \hline
    \hline
  \end{tabular}
  
  \label{tab6}
\end{table}



\end{document}